\documentclass[]{aa}
\usepackage{graphicx,hyperref,txfonts}
\usepackage{subcaption}
\usepackage{array}
\newcommand{\com}[1]{\textnormal{#1}}
\newcommand{\comm}[1]{\textnormal{#1}}
\newcounter{savecounter} % A variable in which we can save the current section
\begin{document}

\title{Model-independent and model-based local lensing properties of CL0024+1654 from multiply-imaged galaxies}
\titlerunning{Model-independent and model-based reconstructions of CL0024}
\author{Jenny Wagner\inst{1}  \and Jori Liesenborgs\inst{2} \and Nicolas Tessore\inst{3}}
\institute{Universit\"at Heidelberg, Zentrum f\"ur Astronomie, Institut f\"ur Theoretische Astrophysik, Philosophenweg 12, 69120 Heidelberg, Germany, Heidelberg Institute for Theoretical Studies, 69118 Heidelberg, Germany \\
\email{j.wagner@uni-heidelberg.de}
\and
Expertisecentrum voor Digitale Media, Universiteit Hasselt, Wetenschapspark 2, B-3590, Diepenbeek, Belgium \\
\email{jori.liesenborgs@uhasselt.be}
\and 
Jodrell Bank Centre for Astrophysics, School of Physics and Astronomy, University of Manchester, Alan Turing Building, Oxford Road, Manchester M13 9PL, UK \\ \email{nicolas.tessore@manchester.co.uk}}
\date{Received XX; accepted XX}

\abstract{\textit{Context.} Local gravitational lensing properties, like convergence and shear, determined at the positions of multiply-imaged background objects, yield valuable information on the smaller-scale lensing matter distribution in the central part of galaxy clusters. Highly distorted multiple images with resolved brightness features like the ones observed in CL0024 allow to study these local lensing properties \comm{and} to tighten the constraints on the properties of dark matter on sub-cluster scale. 
\\
\textit{Aims.} We investigate to which precision local magnification ratios, $\mathcal{J}$, ratios of convergences, $f$, and reduced shears, $g = (g_{1}, g_{2})$, can be determined model-independently for the five resolved multiple images of the source at $z_\mathrm{s}=1.675$ in CL0024. We also determine if a comparison to the respective results obtained by the parametric modelling \comm{tool} Lenstool and by the non-parametric modelling \comm{tool} Grale can detect biases in the lens models. For these model-based approaches we additionally analyse the influence of the number and location of the constraints from multiple images on the local lens properties determined at the positions of the five multiple images of the source at $z_\mathrm{s}=1.675$.
\\
\textit{Methods.} Our model-independent approach uses a linear mapping between the five resolved multiple images to determine the magnification ratios, ratios of convergences, and reduced shears at their positions.With constraints from up to six multiple image systems, we generate Lenstool and Grale models using the same image positions, cosmological parameters, and number of generated convergence and shear maps to determine the local values of $\mathcal{J}$, $f$, and $g$ at the same positions across all methods. 
\\
\textit{Results.} All approaches show \comm{strong} agreement on the local values of $\mathcal{J}$, $f$, and $g$. We find that Lenstool obtains the tightest confidence bounds even for convergences around one using constraints from six multiple image systems, while the best Grale model is generated only using constraints from all multiple images with resolved brightness features and adding limited small-scale mass corrections. Yet, confidence bounds as large as the values themselves can occur for convergences close to one in all approaches. 
\\
\textit{Conclusions.} Our results are in agreement with previous findings, supporting the light-traces-mass assumption and the merger hypothesis for CL0024. Comparing the three different approaches allows to detect modelling biases. Given that the lens properties remain approximately constant over the extension of the image areas covered by the resolvable brightness features, the model-independent approach determines the local lens properties to a comparable precision but within less than a second.

%i%n the $g$ estimates of Lenstool and to check for overfitting by Grale. The results agree to the light-traces-mass assumption and the hypothesis that CL0024 is a merger event.

%do not allow a robust and tight model-independent local lens characterisation. Avoiding overfitting by comparing the local $f$s and $g$s to the model-independent results and tightening their confidence bounds by multiple constraints close to the point at which they are to be determined, we find that Grale is the best modelling ansatz to determine small-scale dark matter properties in CL0024. Hence, local constraints put the most reliable constraints on local lens properties. For cases with resolved features that are spread more uniformly than in CL0024 close to the position of interest, the model-independent approach alone can lead to results of comparable quality as current model-based local $f$s and $g$s. 
}
\maketitle

%%%%%%%%%%%%%%%%%
\section{Motivation and related work}
\label{sec:introduction}

The Frontier Fields Lens Modelling Comparison Project, \cite{bib:Meneghetti2016}, is the most encompassing systematic comparison of lens modelling approaches. It employs simulated data of unresolved multiple images in two artificially generated galaxy clusters.  From this comparison we know that the mass enclosed in the critical curves of a galaxy cluster is determined to only a few percent inaccuracy and imprecision by any lens modelling approach. In contrast, the accuracy and precision of local convergence and shear values strongly depend on the number and positions of the multiple images observed. This, in turn, sets the limits on the accuracy and precision up to which the distribution and properties of small-scale compact dark matter can be determined, e.g.\ in \cite{bib:Diego2017}.

\comm{Simulations of multiple images showing distinctive features in their brightness distributions are still to be developed. So far, real galaxies extracted from deep HST observations have been employed to model the brightness distributions of sources in simulations, as e.g.\ in \cite{bib:Meneghetti2016}. Hence, the multiple images appear unstructured in current simulations because the high magnification regions in the strong lensing regime would require a resolution beyond the one of the HST cameras for the sources to obtain images with resolved brightness features. Thus,} we rely on the galaxy cluster CL0024 to compare the model-independent reconstruction of local ratios of convergences and reduced shear values developed in \cite{bib:Tessore1} with the parametric lens modelling approach Lenstool, and the non-parametric lens modelling approach Grale. The model-independent approach was \com{tested} \comm{using simulations} in \cite{bib:Wagner2}, Lenstool was developped in \cite{bib:Kneib, bib:Jullo2007}, and Grale in \cite{bib:Liesenborgs2006, bib:Liesenborgs2010}.

CL0024 is a well-studied cluster, whose strong lensing properties have been investigated e.g.\ in \cite{bib:Broadhurst, bib:Colley, bib:Jee, bib:Liesenborgs2008, bib:Richard, bib:Umetsu, bib:Zitrin}. Therefore, we do not develop more advanced models but focus on the model comparison with the model-independent approach. The cluster contains one system of five multiple images. Each of them shows six distinctive features in the brightness distribution which can be used to model-independently determine ratios of convergences and reduced shears at the positions of the multiple images, \comm{as described in \cite{bib:Tessore1} and \cite{bib:Wagner2}}. In addition, a multitude of systems of non-resolved multiple images are proposed in this cluster, e.g.\ in \cite{bib:Zitrin} which we will also employ to set up the Lenstool and Grale lens models. 

Using the same cosmological parameter values and positions of multiple image systems, we calculate the Lenstool and Grale lens models. We retrieve the local ratios of convergences and reduced shears from the respective maps at the same positions to compare them to the same lens properties calculated from the model-independent approach at these positions. This direct comparison between the two model-based and the model-independent local lens properties allows to investigate differences and similarities between the different lens reconstruction ansatzes. Furthermore, we investigate whether the comparison can determine, if the light-traces-mass assumption usually employed in parametric lens modelling is fulfilled
% it cannot be rejected as non-overlapping confidence intervals can also imply higher order effects that are not taken into account in our linear transformation approach!
and whether we can set a scale below which further refinements of a non-parametric lens model may overfit the model to the data. This may generate dark matter \comm{artefacts}, as possibly found in \cite{bib:Jee} and discussed in \cite{bib:Ponente}. 

We also investigate the robustness of the local convergence ratios and reduced shear values of the model-based approaches when we reduce the number of multiple image systems that are used for the lens modelling. This analysis shows how strong constraints from multiple images of one system influence the convergence ratios and reduced shear values at the positions of neighbouring multiple images of other systems and whether the multiple image positions of one system suffice as input constraints for the lens model in order to yield the local convergence ratios and reduced shears at these positions. 

The paper is organised as follows: After the introduction of the multiple image systems in CL0024 that we employ in Section~\ref{sec:data}, we describe how the model-independent information is calculated in Section~\ref{sec:mi_information}. Then, we generate the Lenstool and Grale model-based reconstructions of CL0024 in Section~\ref{sec:models} and extract the same information from their convergence and shear maps as can be retrieved from the model-independent approach. Based on this information, a comparison of all three approaches is performed in Section~\ref{sec:comparison} whose results are summarised in Section~\ref{sec:conclusion}.

%%%%%%%%%%%%%%%%%
\section{Multiple image systems in CL0024}
\label{sec:data}

In CL0024, a multitude of multiple image system candidates has been detected and used to reconstruct the lensing properties of the galaxy cluster (see Section~\ref{sec:introduction}). However, only for one five image system a spectroscopic redshift has been measured so far, \cite{bib:Broadhurst}, corroborating the lensing hypothesis for these images. The most systematic collection of multiple image system candidates together with photometric redshift estimates was published in \cite{bib:Zitrin}, which we will use in the following. As none of the candidate systems~2--11 of \cite{bib:Zitrin} is confirmed spectroscopically, not all of them need to be true multiple image systems. Therefore, we employ only a subset of those candidate systems close to system~1, guided by an initial Lenstool reconstruction as detailed in Section~\ref{sec:Lenstool}.

To make the reconstructed local lens properties as comparable as possible, the model-based approaches employ the same positions of the six multiple image systems listed in Table~\ref{tab:multiple_image_systems}. In addition to the points in system~1 listed in Table~\ref{tab:multiple_image_systems}, the model-independent reconstruction and Grale use the positions of up to five additional identifiable bright spots in all images of system~1 listed in Table~\ref{tab:system1}. These points were determined by \com{eye} of the HST ACS/WFC image in the F475W band (PI:Ford 2004\footnote{Based on observations made with the NASA/ESA Hubble Space Telescope, and obtained from the Hubble Legacy Archive, which is a collaboration between the Space Telescope Science Institute (STScI/NASA), the Space Telescope European Coordinating Facility (ST-ECF/ESA) and the Canadian Astronomy Data Centre (CADC/NRC/CSA).}), in which these resolved features are most prominent.

\begin{table}[t]
 \caption{Systems of multiple images employed in the model-based lens reconstructions. The coordinates for system 1 are determined by visual inspection, this system is the only one with a spectroscopic redshift, the properties of the remaining systems are determined as described in \cite{bib:Zitrin} with a BPZ photometric redshift estimation according to \cite{bib:Benitez2000, bib:Benitez2004, bib:Coe}. For comparison with \cite{bib:Zitrin}, we keep their nomenclature (except for system~1) and also list the redshifts $z_\mathrm{m}$ determined by their model.}
\label{tab:multiple_image_systems}
\begin{center}
\begin{tabular}{ccccc}
\hline
\noalign{\smallskip}
  Image & RA $\left[\text{deg}\right]$ & DEC $\left[\text{deg}\right]$ & $z$ & $z_\mathrm{m}$ \\
\noalign{\smallskip}
\hline
\noalign{\smallskip}
$ 1.1$ & $6.65528$ & $17.15463$ & $1.675$                &                        \smallskip \\
$ 1.2$ & $6.65689$ & $17.15678$ & $1.675$                &                        \smallskip \\
$ 1.3$ & $6.65832$ & $17.16154$ & $1.675$                &                        \smallskip \\
$ 1.4$ & $6.64303$ & $17.16495$ & $1.675$                &                        \smallskip \\
$ 1.5$ & $6.64724$ & $17.16173$ & $1.675$                &                        \smallskip \\
\hline
\noalign{\smallskip}
$ 3.1$ & $6.65358$ & $17.15675$ & $2.76^{+0.37}_{-2.59}$ & $2.55^{+0.45}_{-0.20}$ \smallskip \\
$ 3.2$ & $6.64858$ & $17.17178$ & $2.48 \pm 0.34$        &                        \smallskip \\
$ 3.3$ & $6.64475$ & $17.17017$ & $2.51^{+0.34}_{-2.19}$ &                        \smallskip \\
$ 3.4$ & $6.63717$ & $17.16294$ & $2.58^{+0.35}_{-2.38}$ &                        \smallskip \\
\hline
\noalign{\smallskip}
$ 4.1$ & $6.64413$ & $17.16169$ & $2.13^{+0.31}_{-0.33}$ & $1.96 \pm 0.20$        \smallskip \\
$ 4.2$ & $6.64479$ & $17.16153$ & $2.30^{+0.34}_{-0.68}$ &                        \smallskip \\
$ 4.3$ & $6.66138$ & $17.16072$ & $2.28^{+0.36}_{-2.07}$ &                        \smallskip \\ 
\hline
\noalign{\smallskip}
$ 5.1$ & $6.63692$ & $17.16092$ & $0.25^{+2.44}_{-0.12}$ & $2.02 \pm 0.20$        \smallskip \\
$ 5.2$ & $6.65317$ & $17.15886$ & $1.58^{+0.65}_{-1.52}$ &                        \smallskip \\
\hline
\noalign{\smallskip}
$ 8.1$ & $6.65158$ & $17.14886$ & $4.09 \pm 0.50$        & $4.03 \pm 0.50$        \smallskip \\
$ 8.2$ & $6.64588$ & $17.16744$ & $4.16^{+0.51}_{-3.62}$ &                        \smallskip \\
\hline
\noalign{\smallskip}
$10.1$ & $6.65071$ & $17.16175$ & $0.75 \pm 0.17$        & $0.96^{+0.23}_{-0.20}$ \smallskip \\
$10.2$ & $6.65046$ & $17.16186$ & $0.58^{+0.16}_{-0.15}$ &                        \smallskip \\
$10.3$ & $6.64021$ & $17.16183$ & $0.85^{+0.31}_{-0.26}$ &                        \smallskip \\ 
\noalign{\smallskip}
\hline
\end{tabular}
\end{center}
\end{table}

\begin{table}[t]
 \caption{Positions of six bright spots, called reference points (RP), identifiable in each of the five images (I) of system 1 from Table~\ref{tab:multiple_image_systems}. Reference point 6 represents system 1 in Table~\ref{tab:multiple_image_systems}.}
 \label{tab:system1}
\begin{center}
\begin{tabular}{cccc}
\hline
\noalign{\smallskip}
  I & RP & RA $\left[\text{deg}\right]$ & DEC $\left[\text{deg}\right]$ \\
\noalign{\smallskip}
\hline
\noalign{\smallskip}
1 & 1 & $6.65452$ & $17.15418$ \smallskip \\
1 & 2 & $6.65486$ & $17.15421$ \smallskip \\
1 & 3 & $6.65495$ & $17.15453$ \smallskip \\
1 & 4 & $6.65505$ & $17.15463$ \smallskip \\
1 & 5 & $6.65520$ & $17.15470$ \smallskip \\
1 & 6 & $6.65528$ & $17.15463$ \smallskip \\
\hline
\noalign{\smallskip}
2 & 1 & $6.65700$ & $17.15722$ \smallskip \\
2 & 2 & $6.65710$ & $17.15702$ \smallskip \\
2 & 3 & $6.65688$ & $17.15700$ \smallskip \\
2 & 4 & $6.65684$ & $17.15695$ \smallskip \\
2 & 5 & $6.65683$ & $17.15685$ \smallskip \\
2 & 6 & $6.65689$ & $17.15678$ \smallskip \\
\hline
\noalign{\smallskip}
3 & 1 & $6.65800$ & $17.16163$ \smallskip \\
3 & 2 & $6.65823$ & $17.16128$ \smallskip \\
3 & 3 & $6.65813$ & $17.16178$ \smallskip \\
3 & 4 & $6.65815$ & $17.16183$ \smallskip \\
3 & 5 & $6.65824$ & $17.16173$ \smallskip \\
3 & 6 & $6.65832$ & $17.16154$ \smallskip \\
\hline
\noalign{\smallskip}
4 & 1 & $6.64345$ & $17.16557$ \smallskip \\
4 & 2 & $6.64336$ & $17.16524$ \smallskip \\
4 & 3 & $6.64311$ & $17.16524$ \smallskip \\
4 & 4 & $6.64304$ & $17.16517$ \smallskip \\
4 & 5 & $6.64298$ & $17.16503$ \smallskip \\
4 & 6 & $6.64303$ & $17.16495$ \smallskip \\
\hline
\noalign{\smallskip}
5 & 1 & $6.64756$ & $17.16184$ \smallskip \\
5 & 2 & $6.64739$ & $17.16187$ \smallskip \\
5 & 3 & $6.64739$ & $17.16175$ \smallskip \\
5 & 4 & $6.64736$ & $17.16173$ \smallskip \\
5 & 5 & $6.64729$ & $17.16171$ \smallskip \\
5 & 6 & $6.64724$ & $17.16173$ \smallskip \\
\noalign{\smallskip}
\hline
\end{tabular}
\end{center}
\end{table}

%%%%%%%%%%%%%%%%%

\section{Model-independent lens reconstruction}
\label{sec:mi_information}

In \cite{bib:Tessore1}, it was shown how properties of strong gravitational lenses can be recovered in a model-independent fashion from the mapping of multiple images onto each other.
Given $n$ multiple images, the \emph{image maps}~$\varphi_i$ are functions that map the arbitrarily-chosen reference image~$1$ onto the multiple image~$i = 2, \dots, n$.
The Jacobian matrix of the image map~$\varphi_i$ is the \emph{relative magnification matrix}~$\tens T_i = \tens A_i^{-1} \, \tens A_1^{}$ between images~$1$ and~$i$, where~$\tens A$ is the usual magnification matrix,
\begin{equation}\label{eq:A}
	\tens A_i = (1 - \kappa_i) \, \begin{pmatrix}
		1 - g_{i,1} & -g_{i,2} \\
		-g_{i,2} & 1 + g_{i,1}
	\end{pmatrix} \;, \quad
	i = 1, \dots, n \;,
\end{equation}
in terms of the convergence~$\kappa$ and the two components of the reduced shear~$g$
\begin{equation}
g_{i,1} = \dfrac{\gamma_{i,1}}{1-\kappa_i} \;, \quad g_{i,2} =  \dfrac{\gamma_{i,2}}{1-\kappa_i} \;, \quad i = 1,...,5\;,
\label{eq:reduced_shear}
\end{equation}
with $\gamma_{i,1}$ and $\gamma_{i,2}$ denoting the shear components for each image $i$, as defined in the usual notation by \cite{bib:SEF} with respect to the RA/Dec coordinate system.
 
Explicitly writing out the entries of~$\tens T_i$, $i=2,\dots, n$,
\begin{gather}
	\tens T_{i,11}
	= \frac{1 - \kappa_1}{1 - \kappa_i} \, \frac{(1 - g_{1,1})(1 + g_{i,1}) - g_{1,2} \, g_{i,2}}{1 - g_{i,1}^2 - g_{i,2}^2} \;, \label{eq:Ti11} \\
	\tens T_{i,12}
	= \frac{1 - \kappa_1}{1 - \kappa_i} \, \frac{(1 + g_{1,1}) \, g_{i,2} - (1 + g_{i,1}) \, g_{1,2}}{1 - g_{i,1}^2 - g_{i,2}^2} \;, \label{eq:Ti12} \\
	\tens T_{i,21}
	= \frac{1 - \kappa_1}{1 - \kappa_i} \, \frac{(1 - g_{1,1}) \, g_{i,2} - (1 - g_{i,1}) \, g_{1,2}}{1 - g_{i,1}^2 - g_{i,2}^2} \;, \label{eq:Ti21} \\
	\tens T_{i,22}
	= \frac{1 - \kappa_1}{1 - \kappa_i} \, \frac{(1 + g_{1,1})(1 - g_{i,1}) - g_{1,2} \, g_{i,2}}{1 - g_{i,1}^2 - g_{i,2}^2} \;, \label{eq:Ti22}
\end{gather}
the reduced shears~$g_{1,j}$ and $g_{i,j}$, $j=1, 2$, are generally recoverable, while the convergences~$\kappa_1$ and~$\kappa_i$ only appear in the form of a \emph{convergence ratio},
\begin{equation}
	f_i = \frac{1 - \kappa_1}{1 - \kappa_i} \quad i=2,\dots, n \;,
	\label{eq:convergence_ratio}
\end{equation}
that scales all components of $\tens T_i$ equally.
Therefore, neither~$\kappa_1$ nor~$\kappa_i$ can be recovered individually (this is also implied by the mass sheet degeneracy), and only the convergence ratios and reduced shears~$g_{1,1}, g_{1,2}, f_2, g_{2,1}, g_{2,2}, \dotsc$ are observable properties of strong gravitational lenses at first order in the image mapping and hence second order in the deflection potential.

To reconstruct the values of~$f$ and~$g$ from a given relative magnification matrix~$\tens T$, it is useful to construct combinations of the entries~\eqref{eq:Ti11}--\eqref{eq:Ti22},
\begin{gather}
	a_i = \tens T_{i,11} - \tens T_{i,22} = 2 f_i \, \frac{g_{i,1} - g_{1,1}}{1 - g_{i,1}^2 - g_{i,2}^2} \;, \label{eq:a_i} \\
	b_i = \tens T_{i,21} + \tens T_{i,12} = 2 f_i \, \frac{g_{i,2} - g_{1,2}}{1 - g_{i,1}^2 - g_{i,2}^2} \;, \label{eq:b_i} \\
	c_i = \tens T_{i,21} - \tens T_{i,12} = 2 f_i \, \frac{g_{i,1} \, g_{1,2} - g_{1,1} \, g_{i,2}}{1 - g_{i,1}^2 - g_{i,2}^2} \;, \label{eq:c_i} \\
	d_i = \tens T_{i,11} + \tens T_{i,22} = 2 f_i \, \frac{1 - g_{1,1} \, g_{i,1} - g_{1,2} \, g_{i,2}}{1 - g_{i,1}^2 - g_{i,2}^2} \;. \label{eq:d_i}
\end{gather}
Note that $c_i$ and $d_i$ are the curl and divergence of the image map, respectively.
When at least three images $1$, $i$, $j$ are observed, the system of equations~\eqref{eq:a_i}--\eqref{eq:d_i} can be solved for the reduced shear of the reference image,
\begin{gather}
	g_{1,1} = \frac{a_i \, c_j - a_j \, c_i}
	               {b_i \, a_j - b_j \, a_i} \;, \label{eq:g01} \\
	g_{1,2} = \frac{b_i \, c_j - b_j \, c_i}
	               {b_i \, a_j - b_j \, a_i} \;, \label{eq:g02}
\end{gather}
as well as the convergence ratio and shear of multiple image~$i$,
\begin{gather}
	f_i~~\, = \frac{2 \det (\tens T_i)}{a_i \, g_{1,1} + b_i \, g_{1,2} + d_i} \;, \label{eq:f_T} \\
	g_{i,1} = \frac{d_i \, g_{1,1} - c_i \, g_{1,2} + a_i}{a_i \, g_{1,1} + b_i \, g_{1,2} + d_i} \;, \label{eq:g1_T} \\
	g_{i,2} = \frac{c_i \, g_{1,1} + d_i \, g_{1,2} + b_i}{a_i \, g_{1,1} + b_i \, g_{1,2} + d_i} \;, \label{eq:g2_T}
\end{gather}
where $\det (\tens T_i)$ is the determinant. The same expressions hold for multiple image~$j$.

\begin{figure*}[ht]
\centering
  \includegraphics[width=0.89\textwidth]{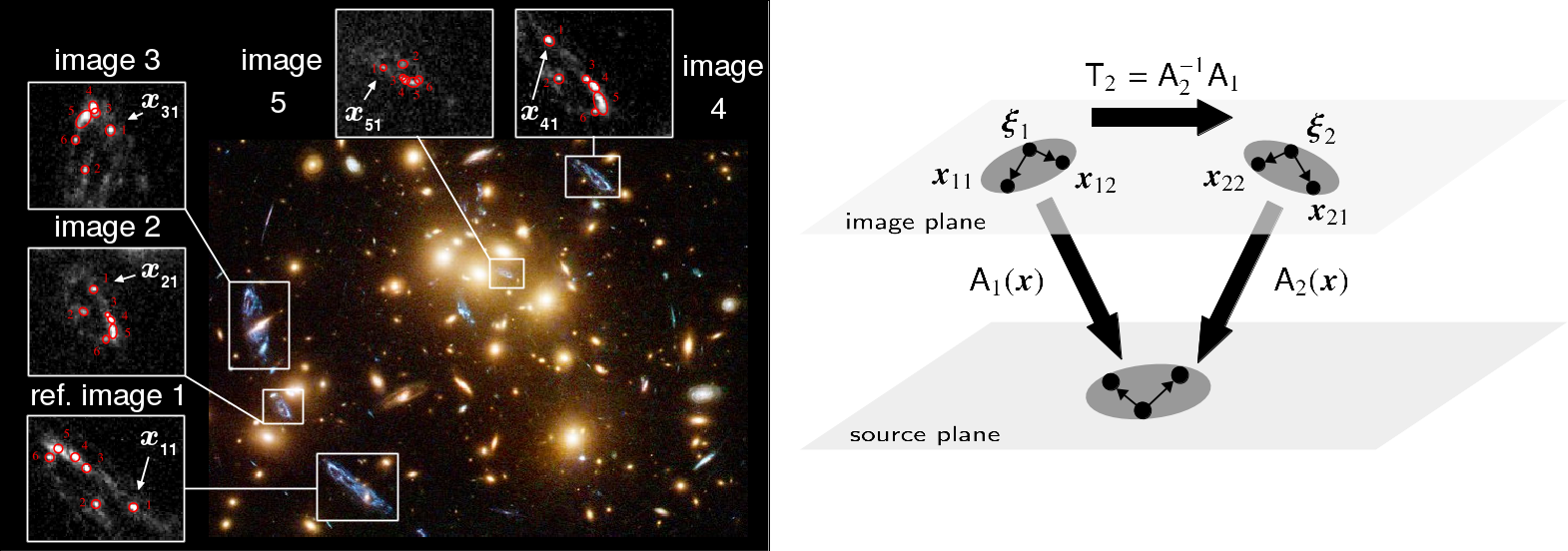}
   \caption{\com{CL0024 with detailed pictures of the five multiple images of system~1 (left). The user-defined reference points in the reference image~1 are marked by red circles, while the red-encircled regions in the remaining images denote 95\% confidence bounds of the locations of the transformed reference points by the best-fit linear transformations. (\textit{Image credits: NASA, ESA, M.~J.~Jee (Johns Hopkins University)}) Visualisation of the principle to extract local lens properties from the linear transformation between multiple images from the same source (right).}}
\label{fig:point_match}
\end{figure*}

 With
\begin{equation}
\mathcal{J}_i \equiv \det(\tens T_i) = \det(\tens A_i)^{-1} \det(\tens A_1) = \frac{\mu_i}{\mu_1} \;, \quad i = 2, \dots, n \;,
\label{eq:J}
\end{equation}
we relate the determinant of the transformation matrix to the magnification ratio between image~$i$, $\mu_i$, and image~1, $\mu_1$, which yields the relative parities between those images and can be compared to the respective, measured flux ratios. Using Equations~\eqref{eq:Ti11} to \eqref{eq:Ti22}, the $\mathcal{J}_i$ can be expressed in terms of the $f_i$, $g_{i,1}$, and $g_{i,2}$, so that they do not yield additional information.

By constructing the image maps~$\varphi_i$, $i=2,\dots, n$, from observations and evaluating their Jacobian matrices, it is possible to measure the relative magnification matrix~$\tens T_i$ directly from data.
This in turn makes it possible to reconstruct the magnification ratios~\eqref{eq:J}, convergence ratios~\eqref{eq:f_T}, and reduced shears~\eqref{eq:g1_T}--\eqref{eq:g2_T} -- i.e.\@ the observable properties -- of the gravitational lens in a completely model-independent manner.

Equations~\eqref{eq:reduced_shear}, \eqref{eq:convergence_ratio}, and \eqref{eq:J} define lens properties that can also be extracted from model-based convergence and shear maps and thus be compared between all three lens reconstruction methods.

\subsection{Linear image mapping by point matching}
\label{sec:ptmatch}

We assume that an observer has found a family of $m$ points,\com{ also called reference points in the following}, individually labelled from~$1$ to~$m$, that reliably show the same features across all multiple images. In addition, we require that convergence and shear do not vary significantly over the area covered by the family of points in each multiple image. Then, it is possible to approximate the image maps as linear transformations of conjugate points between the multiple images. 

Let~$\vec x_{ij}$ denote the point with index~$j=1,\dots,m$ in multiple image $i =2,\dots,n$, and~$\vec x_{1j}$ the corresponding point in the reference image~$1$.
If the assumption of linearity holds, the image mapping is described by a matrix, which in this case is the relative magnification matrix $\tens T_i$.
Mapping the observed points $\vec x_{11}, \vec x_{12}, \dots$ from reference image~$1$ to multiple image~$i$ should recover the observed points $\vec x_{ij}$,
\begin{equation}\label{eq:ptmatch}
	\vec x_{ij} - \vec \xi_i = \tens T_i \, (\vec x_{1j} - \vec \xi_1) \;,
\end{equation}
where $\vec \xi_i$, $\vec \xi_1$ are \textit{anchor points} for the affine transformation between multiple images~$1$ and~$i$.
Lens reconstruction by point matching means finding a relative magnification matrix~$\tens T_i$ that solves this equation for all points $j$ in a given image~$i$.

As the reconstructed $\tens T_i$ is the solution of the linearised mapping over the entire area covered by the family of points in a multiple image, it is not necessarily the same as the relative magnification matrix that would be obtained from a fully non-linear image map at the points $\vec \xi_i$ and $\vec \xi_1$.
The anchor point $\vec \xi_1$ of the reference image is arbitrary, since it can be absorbed into the left-hand side of the equation~\eqref{eq:ptmatch}.
However, it makes sense to pick $\vec \xi_1$ within the observed image, for example as the centroid of the points $\vec x_{1j}$.
The locations of the remaining anchor points $\vec \xi_2, \vec \xi_3, \dotsc$, which are additional free parameters of the reconstruction, can then be understood as the images of $\vec \xi_1$ under the linearised image mapping.
 
In observations, the images~$\vec x_{ij}$ of the reference points~$\vec x_{1j}$ will be localised with some level of uncertainty.
The difference between an observed position~$\vec x_{ij}$ and the prediction~\eqref{eq:ptmatch} can be modelled as a bivariate normal random variable,
\begin{equation}\label{eq:delta}
	\vec \Delta_{ij}
	= (\vec x_{ij} - \vec \xi_i) - \tens T_i \, (\vec x_{1j} - \vec \xi_1)
	\sim \mathcal{N}(\vec 0, \tens \Sigma_{ij}) \;,
\end{equation}
where the uncertainty in the observed position~$\vec x_{ij}$ is described by a covariance matrix $\tens \Sigma_{ij}$.
No uncertainty is associated with the reference points~$\vec x_{1j}$, which are fixed by the observer.
For given relative magnification matrices~$\tens T_i$, the quality of the reconstruction is then quantified by the $\chi^2$-value,
\begin{equation}\label{eq:chi2}
	\chi^2 = \sum_{i=1}^{n-1} \sum_{j=1}^{m} \vec \Delta_{ij}^\top \, \tens \Sigma_{ij}^{-1} \, \vec \Delta_{ij}^{} \;,
\end{equation}
where the sum extends over $n-1$ multiple images of the reference image~$1$ and their $m$ observed points.
Minimising the $\chi^2$-term~\eqref{eq:chi2} returns best-fit values for the relative magnification matrices $\tens T_i$ and anchor points $\vec \xi_i$, which are the degrees of freedom of the reconstruction. \com{Figure~\ref{fig:point_match} visualises the point matching using the six reference points in system~1 of Table~\ref{tab:system1} as example (left) and it shows a schematic how the linear transformation is derived from the magnification matrices of two images with two reference points and one anchor point in each image (right).}

\subsection{Parametrisation of the matrices}

When more than three multiple images of a source are observed, the system of constraints for the convergence ratios and reduced shears is overdetermined, since the $n - 1$ relative magnification matrices have $4n - 4$ coefficients, for $3n - 1$ lens quantities, \citep{bib:Tessore1}.
In this case, the relative magnification matrices~$\tens T_i$ cannot directly be used as the parameters of the reconstruction:
Each pair $i, j$ of images could yield a different reduced shear over the reference image~\eqref{eq:g01}--\eqref{eq:g02}, leading to an inconsistent reconstruction.

To circumvent the problem, a suitable parametrisation of the matrices~$\tens T_i$ can be adopted.
A natural choice are the convergence ratio~$f_i$ and reduced shear components~$g_{i,1}, g_{i,2}$, so that the relative magnification matrices are given by expressions~\eqref{eq:Ti11}--\eqref{eq:Ti22}.
In practice, this leads to a numerically difficult reconstruction:
Due to the non-linear form of the expressions, the parameters $f_i$, $g_{i,1}$, $g_{i,2}$ are strongly correlated, which makes the exploration of the parameter space difficult with simple numerical methods.

%It is therefore necessary to find a suitable parametrisation for the relative magnification matrices \textbf{to consistently recover $g_{1,1}$ and $g_{1,2}$}.
%The straightforward choice is to express the relative magnification matrix~$\tens T_i$ in terms of the convergence ratio~$f_i$ and the reduced shear~$g_i$, as in the original definition~\eqref{eq:Ti11}--\eqref{eq:Ti22}.
%Yet, the highly non-linear form of the entries of~$\tens T_i$ strongly correlates the parameters $f_i$, $g_{i,1}$, $g_{i,2}$, which can quickly lead to a numerically difficult reconstruction.

A more practical parametrisation keeps the optimisation problem as nearly linear as possible.
This can be achieved by noting that for every relative magnification matrix $\tens T_i$, there exists a relation between the $a_i$, $b_i$ and $c_i$ coefficients~\eqref{eq:a_i}--\eqref{eq:c_i} and the reduced shear $g_{1,1}, g_{1,2}$ over the reference image,
\begin{equation}\label{eq:c_i-fix}
	g_{1,2} \, a_i - g_{1,1} \, b_i = c_i \;.
\end{equation}
Hence it is possible to use the coefficients $a_i, b_i, d_i$ for multiple images~$i = 2, \dots, n$, together with the reduced shear~$g_{1,1}, g_{1,2}$ over the reference image, as the parameters of the reconstruction, and write the relative magnification matrices as
\begin{equation}
	\tens T_i = \frac{1}{2} \begin{pmatrix}
		d_i + a_i &
		b_i - c_i \\
		b_i + c_i &
		d_i - a_i
	\end{pmatrix} \;,
\end{equation}
where the coefficient~$c_i$ must be computed from the relation~\eqref{eq:c_i-fix}.
With this parametrisation, the reconstruction remains consistent and, at the same time, easy to handle with standard numerical methods.

\subsection{Implementation}
\label{sec:implementation}

An implementation of the image mapping technique presented here is publicly available.\footnote{%
\url{https://github.com/ntessore/imagemap}}
The \texttt{ptmatch} routine will take a list of reference points with optional uncertainties and perform the point matching described above.
Also provided are converters between relative magnification matrices and lens quantities, as well as utilities for producing mapped images and a source reconstruction.
% a source reconstruction if this is cut out...

The C implementation of the MPFIT routine \cite{bib:Markwardt} is used to minimise the $\chi^2$- term~\eqref{eq:chi2} and returns the best-fit values for the reduced shear $g_{1,1}, g_{1,2}$ of the reference image, the coefficients $a_i, b_i, d_i$ of the relative magnification matrices, and the components $\xi_{i,1}, \xi_{i,2}$ of the anchor points, resulting in a total number of $5n - 3$ parameters.
The number of constraints from $m$ points in $n - 1$ multiple images is $2 \, m \, (n - 1)$.
Hence a minimum of 3 points in 3 images is necessary, in which case the system is solvable, as expected \citep{bib:Tessore1}.

MPFIT requires the $\chi^2$-term~\eqref{eq:chi2} to be the sum of squares of uncorrelated random deviates~$\hat{x}_i$ of unit variance,
\begin{equation}\label{eq:chi2-num}
	\chi^2 = \sum_{i} \hat{x}_i^2 \;.
\end{equation}
To bring the difference terms~\eqref{eq:delta} into the required form, a whitening transform $\tens W_{ij}$ with $\tens W_{ij}^\top \, \tens W_{ij}^{} = \tens \Sigma_{ij}^{-1}$ is applied to the random variables,
\begin{equation}
	\hat{\vec x}_{ij}
	= \tens W_{ij} \, \vec \Delta_{ij}
	= \tens W_{ij} \, (\vec x_{ij} - \vec \xi_i) - \tens W_{ij} \, \tens T_i \, (\vec x_{1j} - \vec \xi_1) \;,
\end{equation}
so that the two components of the result are uncorrelated with unit variance.
For a covariance matrix~$\tens \Sigma$ with variances $\sigma_1^2$, $\sigma_2^2$ and correlation coefficient $\rho$, a possible whitening transform is given by the Cholesky decomposition of the inverse matrix~$\tens \Sigma^{-1}$,
\begin{equation}
	\tens W = \begin{pmatrix}
		\frac{1}{\sqrt{1 - \rho^2} \, \sigma_1} &
		\frac{-\rho}{\sqrt{1 - \rho^2} \, \sigma_2} \\
		0 &
		\frac{1}{\sigma_2}
	\end{pmatrix} \;.
\end{equation}
It is clear that the $\chi^2$-value of the transformed variables,
\begin{equation}\label{eq:chi2-w}
	\chi^2 = \sum_{i=1}^{n-1} \sum_{j=1}^{m} \hat{\vec x}_{ij}^\top \, \hat{\vec x}_{ij}^{} \;,
\end{equation}
is formally the same as the original term~\eqref{eq:chi2}, and at the same time of the required form~\eqref{eq:chi2-num} for MPFIT.

\begin{figure*}[p]
\centering
  \includegraphics[width=0.82\textwidth]{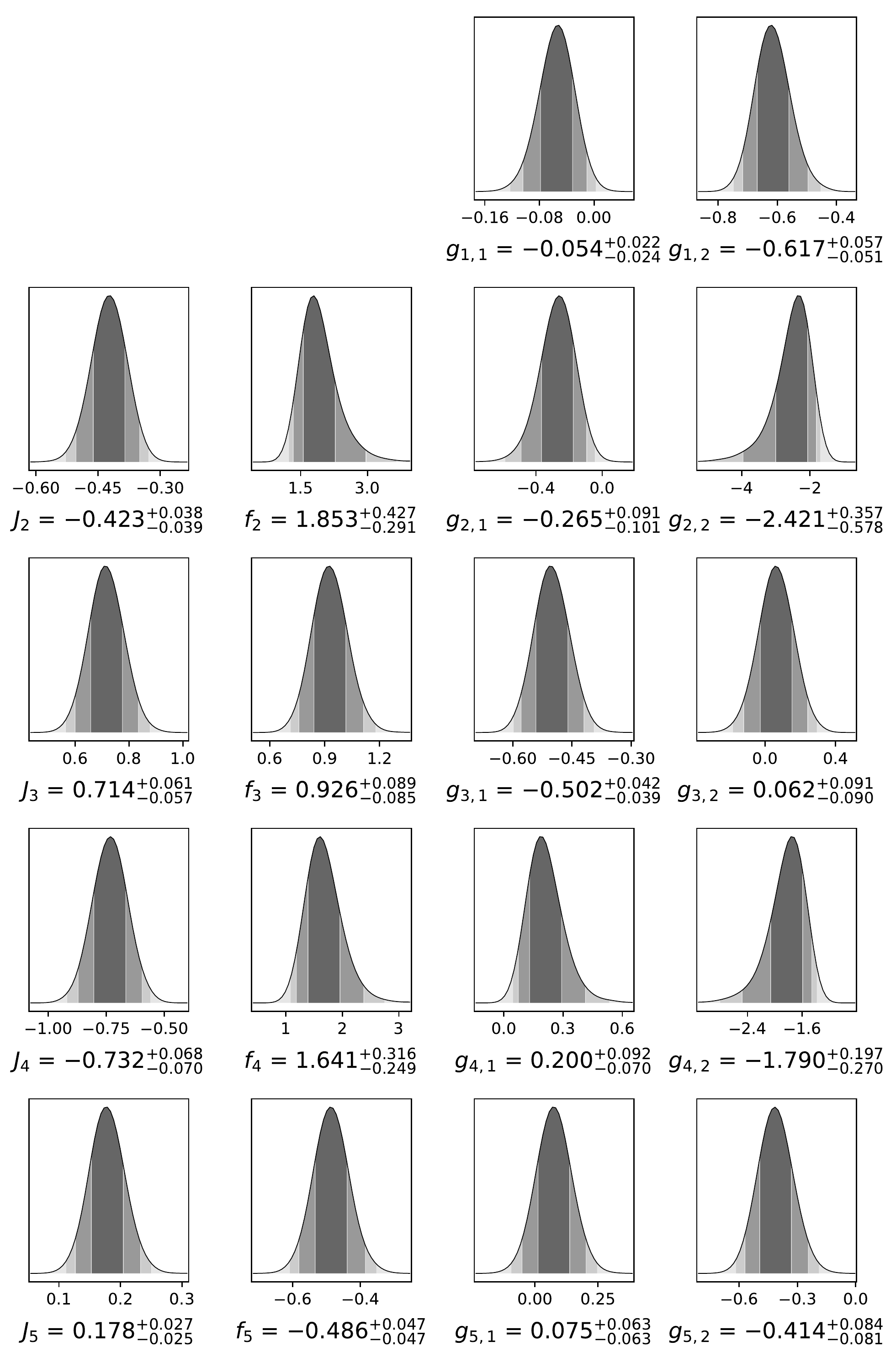}
   \caption{Likelihood distributions of the model-independently determined $\mathcal{J}_i, f_i$, $g_{i,1}$, and $g_{i,2}$, $i=1,...,5$ using all six reference points of Table~\ref{tab:system1}. Dark grey shaded areas mark the region between the 16th and 84th percentile.}
\label{fig:MI_RP6}
\end{figure*}

\begin{figure*}[p]
\centering
  \includegraphics[width=0.82\textwidth]{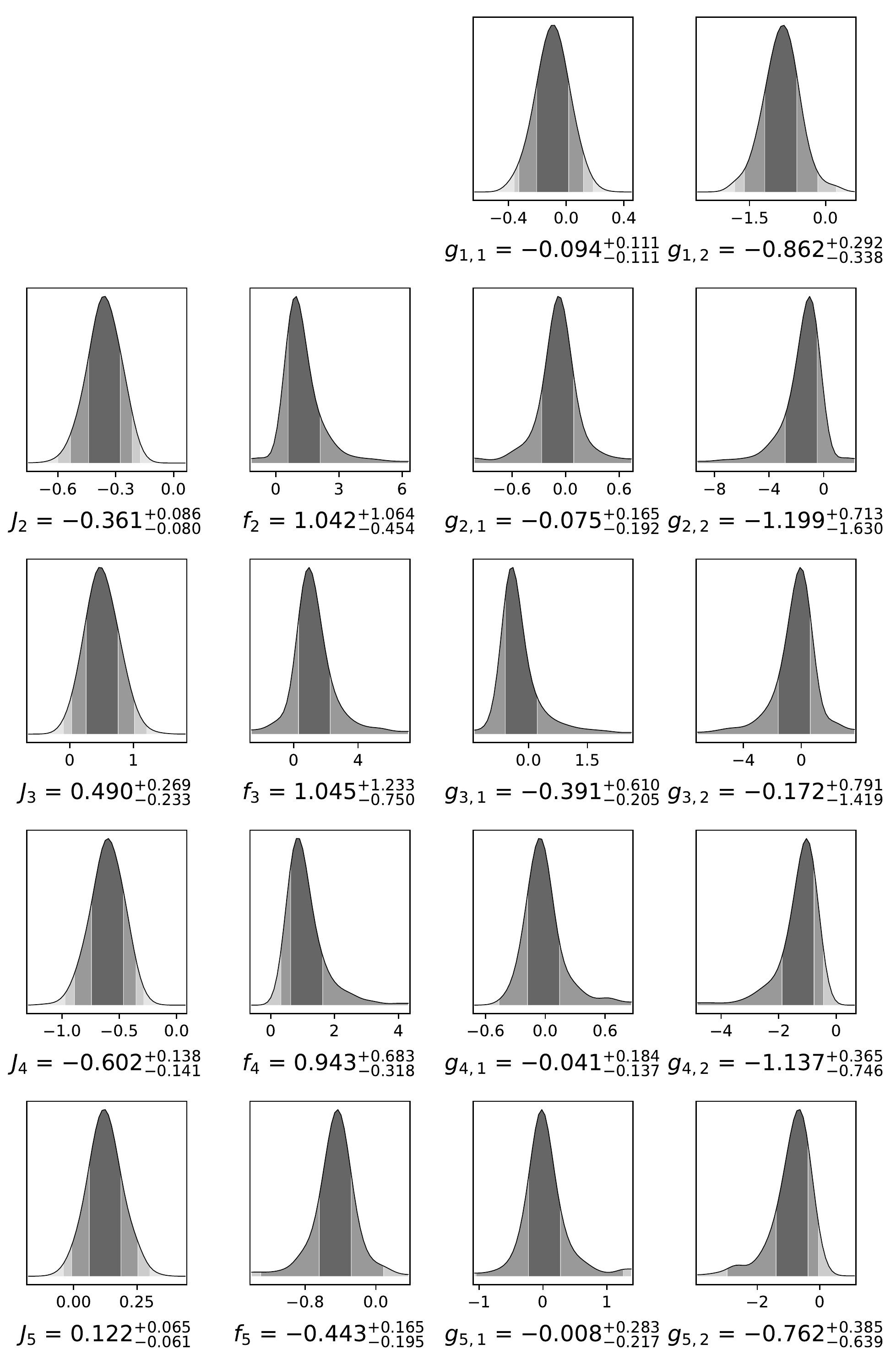}
   \caption{Likelihood distributions of the model-independently determined $\mathcal{J}_i, f_i$, $g_{i,1}$, and $g_{i,2}$, $i=1,...,5$ using the last four reference points of Table~\ref{tab:system1}.  Dark grey shaded areas mark the region between the 16th and 84th percentile.}
\label{fig:MI_RP4}
\end{figure*}

Besides the best-fit values, MPFIT also allows to estimate the covariance matrix of the parameters near the minimum by numerical differentiation.\footnote{%
The numerical differentiation is used for the second derivatives of the $\chi^2$-term.
First derivatives with respect to the parameters are given analytically.
}
Both results can be used together to sample the parameter space using importance sampling from a normal distribution with the estimated covariance matrix and centred on the maximum-likelihood parameters.
This process yields a full likelihood distribution and allows the estimation of robust confidence bounds for the reconstructed lens quantities. As examples, Figures~\ref{fig:MI_RP6} and \ref{fig:MI_RP4} show these likelihood distributions for the two model-independent reconstructions of $\mathcal{J}, f$, and $g$ using \com{all six and the last four reference points of Table~\ref{tab:system1}}, respectively. 

%%%%%
\subsection{\com{Testing and application to CL0024}}

For lensing by a simulated singular isothermal, elliptical lens, \cite{bib:Wagner2} showed that this approach becomes inaccurate when the reference points are spread over distances of 10\% of the distance between the closest multiple images. Then, the prerequisite that convergence and shear variations over the image areas are negligible breaks down. Comparing the spread of the six reference points for the first image around their centre, which is about $1''$, with the distance of this image to the second one, which is ca. $10''$, we observe that this limit may be reached for CL0024. 

Therefore,  we perform one reconstruction of the local lens properties using all six reference points in each image and compare it to the results obtained by discarding the first two reference points in each image. While the former reconstruction may become inaccurate due to the large area covered by the reference points, we reduce the area over which the reference points are spread in the latter reconstruction \com{(see Figure~\ref{fig:point_match} (left))}.

As the approach assumes that variations of convergence and shear are negligible over the area covered by the reference points, all image points in the convex hull of the reference points are assigned the same $f$-, $g$-, \com{and $\mathcal{J}$-}values determined by the linear transformation between the multiple images.

% evaluation
Using all six reference points for all five images listed in Table~\ref{tab:system1}, we obtain the mean $f_i$-, $g_{i,1}$-, and $g_{i,2}$-values and their standard deviation from 10 000 samples as shown in Table~\ref{tab:all_results} in the nineth column block. Discarding the first two reference points and repeating the evaluation for another 10 000 samples, the results are listed in the tenth column block of Table~\ref{tab:all_results}. The likelihood distributions with the median values and the confidence levels from the 16th and 84th percentile determined from the 10~000 samples are shown in Figures~\ref{fig:MI_RP6} and \ref{fig:MI_RP4}.

% reconstruction of the critical curves?

%%%%%%%%%%%%%%%%%
\section{Lens models for CL0024}
\label{sec:models}

To investigate the influence of the number of strong lensing contraints on the local magnification ratios, ratios of convergences, and reduced shear values (Equations~\eqref{eq:convergence_ratio} to \eqref{eq:J}) for the parametric and non-parametric lens modelling methods, we generate lens models 
\begin{itemize}
\item using all six systems of multiple images (as selected according to Section~\ref{sec:selection_of_images}) with Grale and Lenstool, 
\item using only the constraints from system~1 of Table~\ref{tab:multiple_image_systems} with Grale and Lenstool, and
\item using the positions of all six resolved features of system~1 as constraints\footnote{An equivalent Lenstool model cannot be generated because the six resolved features do not suffice to constrain a global large-scale halo and smaller-scale local substructures of dark matter close to the positions of the resolved features.} with Grale. 
\end{itemize}
For the last configuration of the list, we generate one Grale model with small-scale mass corrections and one without, in order to also investigate possible overfitting to the multiple images at the cost of unrealistic mass distributions, as found in \cite{bib:Ponente} for another non-parametric lens modelling approach.

As cosmological parameters, both lens modelling methods use
\begin{equation}
H_0 =  67.80 \dfrac{\mbox{km/s}}{\mbox{Mpc}}\;, \quad  \Omega_{\mathrm{m}} = 0.308 \;, \quad \Omega_\Lambda= 0.692 
\label{eq:cosmological_parameters}
\end{equation}
for the Hubble constant, the matter and dark energy density parameters in agreement with the Planck measurements, \cite{bib:Planck2015}.

% ``blinding''
To ensure that all models have the least human-induced bias possible, the parametric and non-parametric models are simultaneously and independently generated by two of the authors based on the multiple image positions determined by \cite{bib:Zitrin} and the positions of the six resolved features in system~1 determined by the third author prior to modelling.

The quality of the lens models can be compared by the root-mean-square deviations (RMSI) between the model-generated images and the observed ones, which are calculated in the same way for both lens model approaches (see Sections~\ref{sec:Lenstool} and \ref{sec:Grale}). 

% extract convergence and shear maps at the user-specified image positions
From the analytical models, we produce convergence and shear maps at the resolution of 0.05'' per pixel and determine the $\mathcal{J}_i$-, $f_i$-, $g_{i,1}$-, and $g_{i,2}$-values, $i=1,...,5$, and their confidence bounds at the positions of system~1 listed in Table~\ref{tab:multiple_image_systems} to compare these local lens properties for all three lens description approaches in Section~\ref{sec:comparison}.

%%%%%%%%%%%%%%%%%
\subsection{Parametric reconstruction by Lenstool}
\label{sec:Lenstool}

% general introduction
Lenstool is a software package that models gravitational lenses as a superposition of smooth, analytical large-scale dark matter halo profiles of specific type previously selected by the user and takes into account the luminous cluster member galaxies with their smaller-scale dark matter halos\footnote{There is a Lenstool version combining parametric and non-parametric lens modelling, \cite{bib:Jullo2009},  which is not considered here, as it is computationally more intensive.}. The parameters of the latter are determined as an ensemble with the same mass profile from the light-traces-mass assumption and the Tully-Fisher and Faber-Jackson scaling relations. 

As further detailed in \cite{bib:Jullo2007}, the optimum lens model for given ranges of parameter values of the predefined dark matter halo profiles, the catalogue of the brightest member galaxies, and the constraints from the systems of multiple images, is obtained by source plane optimisation or image plane optimisation. %While the source plane optimisation minimises the distance of the back-traced multiple images to the barycentre of the overlap between these back-traced images in the source plane, the image plane optimisation is much more time-consuming and minimises the deviation of the model-generated multiple images to the observed ones in the image plane, projecting the back-traced images from the source plane to the image plane again. 

% potential + brightest galaxies
For all Lenstool lens models, we choose the pseudo-isothermal mass distribution (PIEMD) as analytic large-scale dark matter halo profile, which is also used for the latest Lenstool reconstruction of CL0024 employing strong lensing constraints performed by \cite{bib:Richard} (see references therein). As catalogue of brightest member galaxies, we use the one from the Lenstool homepage\footnote{\url{https://projets.lam.fr/projects/lenstool/wiki}}, also employed in \cite{bib:Richard}. 

% goodness of fit
We assess the quality of our Lenstool models by three goodness-of-fit estimators, as described in \cite{bib:Jullo2007}: the RMSI between the model-generated multiple images and the observed ones, the logarithm of the evidence for that model, $\log(\mathcal{E})$, and its $\chi^2$-value. 

% confidence intervals 
The parameters of the optimum lens model and their confidence bounds are determined by a Bayesian Markov-Chain-Monte-Carlo (MCMC) approach, as detailed in \cite{bib:Jullo2007}. %, for a given set of multiple images, a predefined number and types of large-scale dark matter halos, and the catalogue of brightest member galaxies. 
In the same manner, the redshifts for all systems not having a spectroscopic redshift are predicted to be compared with the measured photometric ones and the ones from the model by \cite{bib:Zitrin}. 

%Computing the convergence and shear maps is not included in the MCMC approach \com{and is separately treated in the post-processing utilities of Lenstool}. 
To obtain magnification ratios, ratios of convergences, and reduced shear values at the positions of system~1 listed in Table~\ref{tab:multiple_image_systems} according to Equations~\eqref{eq:convergence_ratio} to \eqref{eq:J} with confidence bounds, we generate 30 \com{convergence and shear maps} and use the average of the retrieved values and their standard deviation for the comparison in Section~\ref{sec:comparison}. 
% In addition, we 10 000 times draw 30 models without replacement from a set of 40 Lenstool models, determine the local lens properties and average over these. 
%The likelihood distributions of the 10 000 averaged $\mathcal{J}_i, f_i, g_{i,1}$, and $g_{i,2}$, $i=1,...,5$, of our Lenstool models are shown in Appendices~\ref{app:LT_6s_histogram} and \ref{app:LT_1s_histogram}.  

%%%%%%%%%%%%%%%%%
\subsubsection{Selection of multiple image systems and number of dark matter halos}
\label{sec:selection_of_images}

We generate Lenstool models with one, two, and three PIEMD large-scale dark matter halos employing all multiple image systems of \cite{bib:Zitrin} and the catalogue of the 85 brightest member galaxies from the Lenstool homepage. Since these models are only used for selection purposes, we optimise them using the fast source plane optimisation. Appendix~\ref{app:lenstool_sp_configuration} shows the configuration file for one PIEMD dark matter halo.  In agreement with \cite{bib:Richard}, we find that the minimal most likely number of dark matter halos is two, taking into account the uncertainty limits of the optimisation procedure. Appendix~\ref{app:selection} shows the critical curves and caustics of the three models, \comm{their goodness-of-fit measures,} and the positions of all multiple images on the HST ACS/WFC image in the F475W band (PI:Ford 2004). 

\comm{With the same three lens models, we select the set of multiple image systems to generate the best-fit Lenstool model by considering the RMSI for all indiviual multiple image sytems and keeping the multiple image systems with the lowest RMSI that sample the vicinity of system~1. Further details about the selection process leading to the set of employed multiple image systems in Table~\ref{tab:multiple_image_systems} can be found in Appendix~\ref{app:selection}.} 

%For the lens models with two and three PIEMDs, the first potential in Appendix~\ref{app:lenstool_sp_configuration} is duplicated and the number of lenses adapted. 

%%%%%%%%%%%%%%%%%
\subsubsection{Reconstruction with six multiple image systems}
\label{sec:lenstool_six_image_systems}

The best-fit Lenstool model is thus calculated using the multiple images of Table~\ref{tab:multiple_image_systems}, two PIEMD large-scale dark matter halo profiles, and the catalogue of the 85 brightest member galaxies. Since the model-independent approach employs image plane observables to retrieve local lens properties at the position of the multiple images, we optimise this model in the image plane. 
Appendix~\ref{app:lenstool_ip_configuration} shows the configuration file to obtain this model. 

Lenstool also solves for the unknown redshifts of the five multiple image systems that have not been analysed spectroscopically. We obtain
\begin{align}
z_3 &= 3.49 \pm 0.39\;, \quad z_4 = 2.04 \pm 0.11\;, \quad z_5 = 1.98 \pm 0.14 \;, \nonumber\\
z_8 &= 4.64 \pm 0.54 \;, \quad z_{10} = 0.83 \pm 0.04
\end{align}
as mean values and standard deviations from the implemented MCMC sampling. 
Comparing these results to the ones obtained by \cite{bib:Zitrin} in Table~\ref{tab:multiple_image_systems}, we observe that they agree within their uncertainty bounds except for sytem~3, which Lenstool estimates much higher. Figure~\ref{fig:lenstool_models} (left) shows the critical curves and caustics that this best-fit Lenstool model produces.

To determine the quality of the set of 30 lens models, as shown in Table~\ref{tab:all_qof}, we determine the average and standard deviation of the total RMSI, $\log(\mathcal{E})$, and $\chi^2$ over all models. In addition, we calculate the RMSI per multiple image system as listed in the second column of Table~\ref{tab:rmsi_comparison}, and observe that all systems show an RMSI lower than 1'', yielding the overall RMSI of 0.6''. Subsequently, we extract the lens properties of Equations~\eqref{eq:reduced_shear}, \eqref{eq:convergence_ratio}, and \eqref{eq:J} from their convergence and shear maps and list the average and standard deviation in the second column block of Table~\ref{tab:all_results}. 
%Figure~\ref{fig:LT_6s} shows the likelihood distributions of 10 000 averaged values of all lens properties generated as detailed in Section~\ref{sec:Lenstool}.

\begin{table}[t]
\caption{Degrees of freedom (DOF), average logarithmic evidence ($\log(\mathcal{E})$), RMSI in arcseconds over all image systems, and total $\chi^2$ and their standard deviations for 30 Lenstool models for CL0024 using the multiple image systems of Table~\ref{tab:multiple_image_systems} as constraints (second row), or only system~1 of Table~\ref{tab:multiple_image_systems} (third row). RMSI in arcseconds over all image systems for the Grale models generated with the multiple image systems of Table~\ref{tab:multiple_image_systems} (fourth row), using system~1 (fifth row), using all points of Table~\ref{tab:system1} (sixth row), and using all points of Table~\ref{tab:system1} including small-scale mass corrections (last row).}
\label{tab:all_qof}
\begin{center}
\begin{tabular}{ccccc}
\hline
\noalign{\smallskip}
Model & DOF & $\log(\mathcal{E})$ & RMSI & $\chi^2$ \\
\noalign{\smallskip}
\hline
\noalign{\smallskip}
Sect.~\ref{sec:lenstool_six_image_systems} & 18 &  $-158 \pm 53$ & $0.60 \pm  0.07$ &  $176 \pm 46$ \\
Sect.~\ref{sec:lenstool_one_image_system} &   0 &  $-\phantom{1}64 \pm 20$ & $0.56 \pm  0.41$ &  $\phantom{1}60 \pm 53$ \\
\noalign{\smallskip}
\hline
\noalign{\smallskip}
Sect.~\ref{sec:grale_six_image_systems} & -- & -- & $0.93 \pm 0.97$ & -- \\
Sect.~\ref{sec:grale_one_image_system} & -- & -- & $0.03 \pm 0.05 $ & -- \\
Sect.~\ref{sec:grale_all_reference_points} & -- & -- & $0.25 \pm 0.09$ & -- \\
Sect.~\ref{sec:grale_all_reference_points_plus_corrections} & -- & -- & $0.13 \pm 0.06$ & -- \\
\noalign{\smallskip}
\hline
\end{tabular}
\end{center}
\end{table}

\begin{table}[t]
\caption{Average RMSI and standard deviation in arcseconds per multiple image system of Table~\ref{tab:multiple_image_systems}, for the Lenstool model of Section~\ref{sec:lenstool_six_image_systems} (second column) and the Grale model of Section~\ref{sec:grale_six_image_systems} (third column) obtained from 30 individual lens models.}
\label{tab:rmsi_comparison}
\begin{center}
\begin{tabular}{ccc}
\hline
\noalign{\smallskip}
System & RMSI & RMSI \\
 & (Lenstool) & (Grale) \\
\noalign{\smallskip}
\hline
\noalign{\smallskip}
  1  &  $0.81 \pm 0.09$ & $0.68 \pm 0.28$ \\ 
  3 &  $0.76 \pm 0.15$  & $1.52 \pm 2.26$ \\
  4 &   $0.32 \pm 0.11$ & $0.09 \pm 0.18$ \\
  5 &  $0.24 \pm 0.13$  & $0.04 \pm 0.04$ \\  
  8 &  $0.32 \pm 0.16$  & $0.02 \pm 0.02$ \\
10 &  $0.42 \pm 0.14$  & $0.41 \pm 0.70$ \\
\noalign{\smallskip}
\hline
\end{tabular}
\end{center}
\end{table}

\com{We employ a constant mass-to-light ratio for the catalogue of brightest cluster member galaxies, as is usually done, see e.g.\ \cite{bib:Meneghetti2016}. In order to test the influence of a non-constant mass-to-light ratio, we change the default slope for the cut radius of the galaxies in the configuration file of Appendix~\ref{app:lenstool_ip_configuration} from 4 to 2.5 to reproduce the fundamental plane (see \cite{bib:Caminha} and references therein). We generate one model by image plane optimisation and employ the \emph{bayesMap}-utility to calculate 30 convergence and shear maps from the MCMC-data of this model.}

\com{Compared to the values listed in the second column of Table~\ref{tab:all_qof}, $\log (\mathcal{E}) = -193$, RMSI $= 0.72$, and $\chi^2 = 246$ indicate that the quality of the model with non-constant mass-to-light ratio is worse. Yet, the estimated redshifts 
\begin{align}
z_3 &= 2.47 \pm 0.81\;, \quad z_4 = 1.94 \pm 0.11\;, \quad z_5 = 1.81 \pm 0.17 \;, \nonumber\\
z_8 &= 3.58 \pm 1.19 \;, \quad z_{10} = 1.00 \pm 0.09
\end{align}
are closer to the ones found in \cite{bib:Zitrin} and also in agreement with the photometric measurements within their ranges of uncertainties. The averages and standard deviations of $\mathcal{J}_i, f_i$, $g_{i,1}$, and $g_{i,2}$, $i=1,...,5$, according to Equations~\eqref{eq:reduced_shear}, \eqref{eq:convergence_ratio},  and \eqref{eq:J} are shown in the third column block of Table~\ref{tab:all_results}.}

%%%%%%%%%%%%%%%%%
\subsubsection{Reconstruction with system~1}
\label{sec:lenstool_one_image_system}

\begin{figure*}[p]
\centering
  \includegraphics[width=0.35\textwidth]{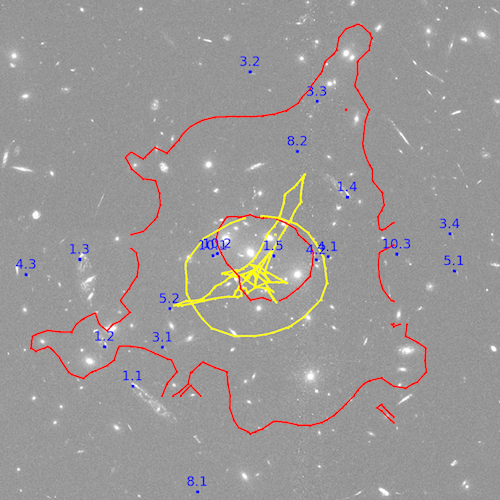}
  \hspace{10ex}
  \includegraphics[width=0.35\textwidth]{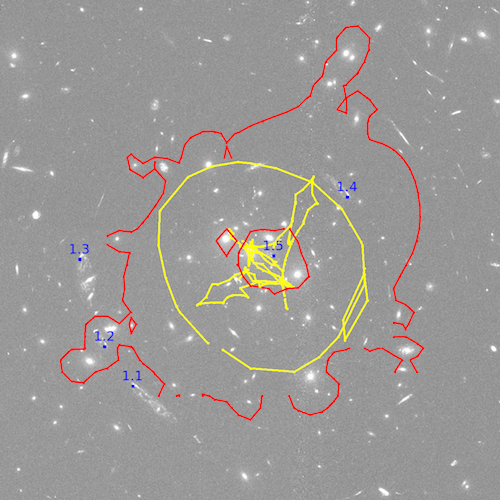}
   \caption{Lenstool models employing the 85 brightest member galaxies and two large-scale PIEMD dark matter halos and all multiple image systems of Table~\ref{tab:multiple_image_systems} detailed in Section~\ref{sec:lenstool_six_image_systems} (left), or only one large-scale PIEMD dark matter halo and system~1 of Table~\ref{tab:multiple_image_systems} detailed in Section~\ref{sec:lenstool_one_image_system} (right). The critical curves for $z_\mathrm{s}=1.675$ determined by the marching squares algorithm (see Appendix~\ref{app:lenstool_ip_configuration}) are marked in red, the caustics in yellow and the multiple images in blue.}
\label{fig:lenstool_models}
\end{figure*}

Reducing the number of strong lensing constraints to system~1 listed in Table~\ref{tab:multiple_image_systems}, the free lens model parameters of only one PIEMD large-scale dark matter halo and of the catalogue of the brightest member galaxies can be determined, if we additionally fix the PIEMD cut radius, chosen to be 1000''. Adapting the configuration file in Appendix ~\ref{app:lenstool_ip_configuration} accordingly, we determine the most likely lens model by image plane optimisation, as in Section~\ref{sec:lenstool_six_image_systems} and show its critical curves and caustics in Figure~\ref{fig:lenstool_models} (right). We generate 30 of these lens models and list the resulting average values of the quality measures and their standard deviations below the ones for the model of Section~\ref{sec:lenstool_six_image_systems} in Table~\ref{tab:all_qof}. Analogously, the $\mathcal{J}_i, f_i$, $g_{i,1}$, and $g_{i,2}$, $i=1,...,5$, according to Equations~\eqref{eq:reduced_shear}, \eqref{eq:convergence_ratio}, and \eqref{eq:J} are shown in the fourth column block of Table~\ref{tab:all_results}. 
%and the likelihood distributions as detailed in Section~\ref{sec:Lenstool} are shown in Figure~\ref{fig:LT_1s}.

%Furthermore, we determine the average velocity dispersion and its standard deviation of the 30 models to be
%\begin{equation}
%\sigma_\mathrm{v} =  (1059 \pm 124 ) \mbox{km/s} \;,
%\end{equation}
%by converting the Lenstool output $v_\mathrm{p}$ 

 \begin{figure*}[p]
\centering
 \includegraphics[width=0.35\textwidth]{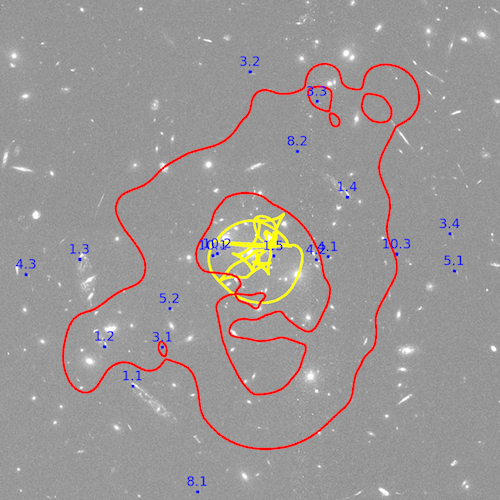}
\hspace{10ex}
  \includegraphics[width=0.35\textwidth]{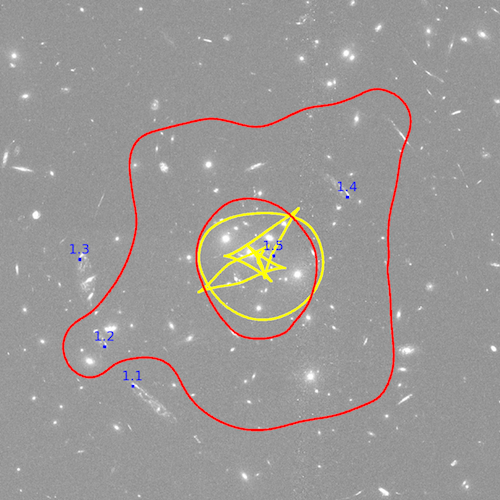}
   \caption{Grale models employing all multiple image systems of Table~\ref{tab:multiple_image_systems} detailed in Section~\ref{sec:grale_six_image_systems} (left), and only system~1 of Table~\ref{tab:multiple_image_systems} detailed in Section~\ref{sec:grale_one_image_system} (right). The critical curves for $z_\mathrm{s}=1.675$ determined by the sign change of the determinant of the magnification matrix are marked in red, the multiple images are marked in blue.}
\label{fig:grale_models}
\end{figure*}

 \begin{figure*}[p]
\centering
 \includegraphics[width=0.35\textwidth]{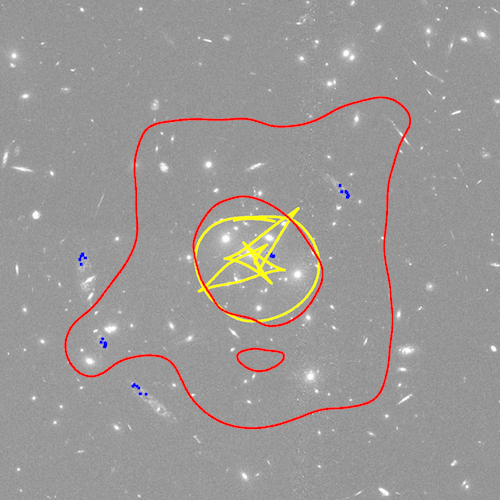}
\hspace{10ex}
  \includegraphics[width=0.35\textwidth]{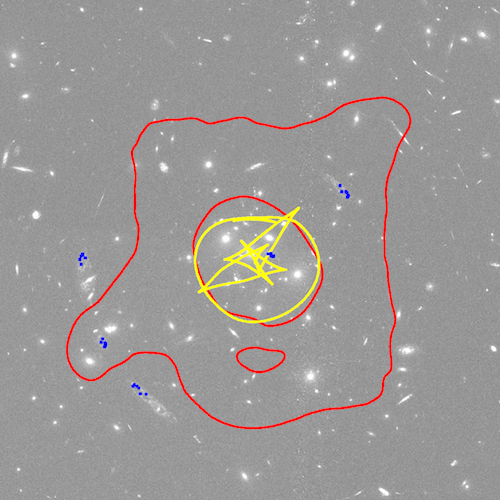}
   \caption{Grale models employing all reference points of Table~\ref{tab:system1} as detailed in Section~\ref{sec:grale_all_reference_points} (left), and including small-scale mass corrections as detailed in Section~\ref{sec:grale_all_reference_points_plus_corrections} (right).The critical curves and caustics are analogous to the ones in Figure~\ref{fig:grale_models}.}
\label{fig:grale_models2}
\end{figure*}

 %%%%%%%%%%%%%%%%%
\subsection{Non-parametric reconstruction by Grale}
\label{sec:Grale}

% general information (also comparison to Lenstool)
Contrary to parametric lens reconstruction techniques like Lenstool, so-called non-parametric or free-form lens reconstruction algorithms make no assumptions about any correlation between the dark and luminous matter distributions. Instead of fitting specific lens models to a given set of multiple image systems and bright cluster member galaxies, they reconstruct the lensing mass distribution in terms of basis functions whose number, locations, and parameters are determined by the constraints the set of multiple image systems provides. For our comparison, we employ Grale by \cite{bib:Liesenborgs2006, bib:Liesenborgs2010}, which divides the region of interest into a uniform grid and assigns a Plummer mass profile, \cite{bib:Plummer}, as basis function to each grid cell. Inspired by the work of \cite{bib:Brewer}, a genetic algorithm determines the weight for each basis function by maximising the overlap of the back-traced images of all sets of multiple images in the source plane. Subdividing the grid in comparably more massive regions, the lens model is refined iteratively until the level of detail is reached which is desired and achievable given the strong lensing constraints.  Hence, similar to Lenstool, the same multiple image positions are employed, which set the scale down to which the dark matter distribution is reconstructed. 

Contrary to Lenstool, this method optimises the lens model only in the source plane and no information on the brightest cluster member galaxies is used. Instead, nullspace information is employed as a further constraint, i.e. the lens model should not generate images in regions not containing any observed ones. 

% evaluating the goodness of fit
As goodness-of-fit measure, the overlap of the back-traced multiple images for all sets of multiple images in the source plane is used.  Furthermore, we check whether a caustic intersects the back-traced multiple images of system~1 and that the model does not produce any further images for the source of system~1. 
%The latter assumption might be incorrect, but the probability for a source to have more than five multiple images is very low. 

Running the genetic algorithm as further detailed in Appendix~\ref{app:grale_configuration} multiple times produces lens models that differ from each other due to the dependency of the optimisation procedure on its initial conditions and lens model degeneracies in sparsely constrained regions. To obtain the best-fit Grale lens model, we average over all of these lens models. The RMSI between the model-generated multiple images and the observed ones is calculated in the same way as done by Lenstool. For our comparison in Section~\ref{sec:comparison}, 30 individual models are used to determine the average and the standard deviation of their magnification ratios, ratios of convergences, and reduced shear maps according to Equations~\eqref{eq:reduced_shear}, \eqref{eq:convergence_ratio}, and \eqref{eq:J} at the positions of system~1 in Table~\ref{tab:multiple_image_systems}. To be consistent with the model-independent approach and Lenstool, we convert the Grale results from the world coordinate system to the pixel coordinates used by the other methods by interchanging the sign of $g_{i,2}$, $i=1,...,5$.

%Analogously to Section~\ref{sec:Lenstool}, we also generate 10 000 Grale models by drawing 30 individual genetic algorithm models without replacement out of 40 and averaging over their lens properties. We plot the likelihood distribution for the lens properties of 10 000 Grale models like the one in Section~\ref{sec:grale_all_reference_points_plus_corrections} as one example in Appendix~\ref{app:G_RPc_histogram}.

%%%%%%%%%%%%%%%%%
\subsubsection{Reconstruction with six multiple image systems}
\label{sec:grale_six_image_systems} % task C for Jori

Contrary to Lenstool, Grale cannot solve for the unknown redshifts of the five multiple image systems without spectroscopic redshift, but requires them as input parameters. As the photometric redshift estimates are subject to high uncertainties and a large scatter between the individual images of one set, we use the well-constrained model-predicted redshifts of \cite{bib:Zitrin}, listed in the right-most column of Table~\ref{tab:multiple_image_systems}. They are in good agreement with the photometric redshifts. \com{The redshifts predicted by our Lenstool models are not employed at this stage in order to keep our lens models independent. After the comparison, we investigate the influence of the Lenstool-predicted redshifts on the Grale modelling in Section~\ref{sec:grale_comparison}.}

As the sources of the multiple image systems lie at different redshifts, the mass sheet degeneracy should be broken sufficiently by the observations, and therefore the weight of a mass-sheet basis function is determined in addition to the weights of the Plummer basis functions. Running Grale 30 times with these specifications, we obtain the mean values of the lens properties and their standard deviations at the positions of system~1 (see Table~\ref{tab:multiple_image_systems}) as listed in the fifth colum block of Table~\ref{tab:all_results}. 

By considering the magnification ratios, ratios of convergences, and reduced shears, any effect due to breaking the mass sheet degeneracy is divided out again. However, including the mass sheet component as an overall mass offset is necessary for the genetic algorithm to achieve the good reconstruction quality shown in Table~\ref{tab:all_qof}. Figure~\ref{fig:grale_models} (left) shows the critical curves of the resulting model, analogously to Figure~\ref{fig:lenstool_models}.

To compare the quality of fit for the Grale models with those of Lenstool, we also list the RMSI for all multiple images in the fourth row of Table~\ref{tab:all_qof} and the RMSI for the individual multiple image systems in Table~\ref{tab:rmsi_comparison}. As can be observed, both approaches are able to reconstruct the multiple images with an overall RMSI below 1'', which is also true for the individual systems except system~3.

%%%%%%%%%%%%%%%%%
\subsubsection{Reconstruction with system~1} 
\label{sec:grale_one_image_system} % task A for Jori

Next, we reduce the six systems of multiple images used in Section~\ref{sec:grale_six_image_systems} to system~1 of Table~\ref{tab:multiple_image_systems}. Now, the mass sheet degeneracy cannot be broken and the additional mass sheet component introduced in the optimisation process for the Grale model of Section~\ref{sec:grale_six_image_systems} is dropped. Using these specifications and running Grale 30 times, we obtain the results at the positions of system~1 (see Table~\ref{tab:multiple_image_systems}) listed in the sixth column block of Table~\ref{tab:all_results} and shown in Figure~\ref{fig:grale_models} (right).

As for the previous Grale model, we also determine the RMSI for system~1 and add it to Table~\ref{tab:all_qof}. With fewer constraints to meet, the RMSI drops significantly, as expected. 

%%%%%%%%%%%%%%%%%
\subsubsection{Reconstruction with all reference points of system~1} 
\label{sec:grale_all_reference_points} % task B for Jori

As Grale is able to selectively refine the resolution of the lens model in more massive regions, we also generate a Grale lens model using all reference points of Table~\ref{tab:system1}, keeping all other specifications as described in the previous section, i.e. without a mass-sheet basis function. The resulting values for the magnification ratios, ratios of convergences, and reduced shears at the positions of system~1 from Table~\ref{tab:multiple_image_systems} are shown in the seventh column block of Table~\ref{tab:all_results} and in Figure~\ref{fig:grale_models2} (left). 

Due to the increasing number of constraints compared to the Grale model of Section~\ref{sec:grale_one_image_system}, the RMSI of all reference points, shown in the sixth row of Table~\ref{tab:all_qof}, increases.

%%%%%%%%%%%%%%%%%
 \begin{figure}[t]
\centering
  \includegraphics[width=0.45\textwidth]{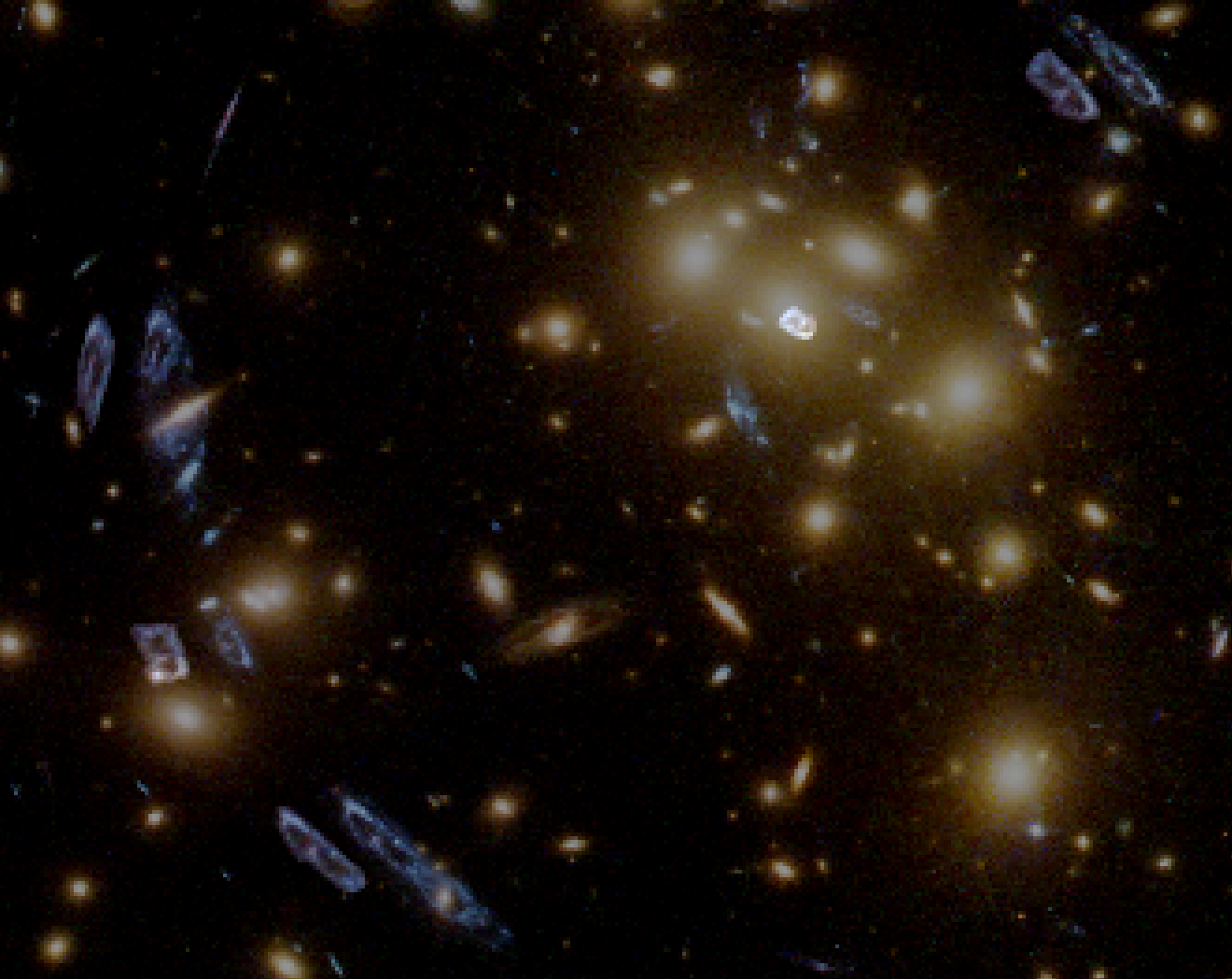}
   \caption{Overlay between the observed multiple images of system~1 in Table~\ref{tab:multiple_image_systems} (\textit{Image credits: NASA, ESA, M.~J.~Jee (Johns Hopkins University)}) and the images generated by the Grale model of Section~\ref{sec:grale_all_reference_points_plus_corrections} using the back-projected first image as source (for visualisation purposes, the model-generated images are displayed with an offset to the left of the observed images). }
\label{fig:grale_relensed}
\end{figure}

\subsubsection{Reconstruction with all reference points of system~1 including small-scale mass corrections} 
\label{sec:grale_all_reference_points_plus_corrections} % task B for Jori

For the last Grale model, we use the same configuration as for Section~\ref{sec:grale_all_reference_points}. But before averaging over all models, we employ a 48 times 48 uniform grid to add small-scale mass corrections to each of the 30 individual solutions generated by the genetic algorithm (see \cite{bib:Liesenborgs2008} for further details). This causes the RMSI to decrease compared to the model of Section~\ref{sec:grale_all_reference_points}, as shown in Table~\ref{tab:all_qof}. In Section~\ref{sec:comparison}, we analyse whether this step introduces unrealistic small-scale dark matter clumps by comparing the local $\mathcal{J}_i, f_i,  g_{i,1}$, and $g_{i,2}$ as listed in the eigth column block of Table~\ref{tab:all_results} to the results from the other approaches. \com{The check for overfitting is necessary, as, for instance, \cite{bib:Ponente} discovered that their algorithm generated unphysical dark matter structures in the lensing mass distribution when forced to optimally match the constraints from the multiple images. Thus, they concluded that the ring-like dark matter structure in CL0024 proposed by \cite{bib:Jee} might be caused by overfitting. To avoid overfitting,} we limit the total amount of small-scale mass corrections to 10\% of the mass already assigned to the cluster. This procedure is able to reproduce the observed images, as shown in Figure~\ref{fig:grale_relensed}, in which we overlay the observed multiple images with the multiple images generated by Grale. The latter are determined by back-projecting the first image of system~1 to the source plane and then mapping this source to the image plane again, using the model discussed in this section. The model-generated images are shown with an offset to the left of the observed ones.  

The critical curves and caustics for this lens model are shown in Figure~\ref{fig:grale_models2} (right). 
%Figure~\ref{fig:G_RPc} shows the histograms of the likelihoods of the local lens properties.

%%%%%%%%%%%%%%%%%
\section{Comparison of the approaches}
\label{sec:comparison}

 The comparison is performed by an automated script after all data have been collected and the evaluation scheme has been defined. As detailed in Sections~\ref{sec:mi_information} and \ref{sec:models}, all results of the model-independent and model-based approaches are summarised in Table~\ref{tab:all_results}. First, we compare the different lens reconstructions for each approach before we compare the reconstructions among the different approaches. 
 %All results listed in Table~\ref{tab:all_results} are visualised in Figure~\ref{fig:all_results}. 
 The lens models are of comparable, good quality, which can be read off Tables~\ref{tab:all_qof} and \ref{tab:rmsi_comparison}, so that effects due to varying quality are negligible. 

% \begin{figure*}[hp!]
%\centering
%  \includegraphics[width=0.475\textwidth]{pictures/f_all_300.png}
%  \hspace{4ex}
%  \includegraphics[width=0.475\textwidth]{pictures/f_all_zoom_300.png}
% \\[12ex]
%   \includegraphics[width=0.475\textwidth]{pictures/g1_all_300.png}
%  \hspace{4ex}
%  \includegraphics[width=0.475\textwidth]{pictures/g1_all_zoom_300.png}
%  \\[12ex]
%   \includegraphics[width=0.475\textwidth]{pictures/g2_all_300.png}
%  \hspace{4ex}
%  \includegraphics[width=0.475\textwidth]{pictures/g2_all_zoom_300.png}
%   \caption{Visualisation of all results listed in Table~\ref{tab:all_results}: $f_i$-values, $i=1,...,5$, according to Equation~\ref{eq:convergence_ratio} for all lens reconstruction approaches for all multiple images of system~1 in Table~\ref{tab:multiple_image_systems} (left) and the same plot with a zoom on the $y$-axis (right) (top row), $g_{i,1}$- and $g_{i,2}$-values according to Equation~\ref{eq:reduced_shear} (left) and their zoomed versions (right) (centre row, bottom row, respectively).}
%\label{fig:all_results}
%\end{figure*}

\begin{sidewaystable*}[hp!]
 \caption{Synopsis of $\mathcal{J}_i, f_i,  g_{i,1}$, and $g_{i,2}$, $i=1,...,5$ of multiple image system~1 of Table~\ref{tab:multiple_image_systems} obtained by a Lenstool reconstruction using all six image systems in Table~\ref{tab:multiple_image_systems} \com{with (LT~6~s) and without (LT~6~s~s) constant mass-to-light ratio of the cluster member galaxies}, and using only the positions of system~1 in Table~\ref{tab:multiple_image_systems} \com{and constant mass-to-light ratio for the member galaxies} (LT~s~1), from Grale reconstructions using the six multiple image systems (G~6~s) or using system~1 of Table~\ref{tab:multiple_image_systems} (G~s~1) and from the Grale model using all points in Table~\ref{tab:system1} without (G RP) and with a final small-scale mass correction (G RP c), and from the model-independent approach using all six reference points (MI~6~RP) or the last four reference points of Table~\ref{tab:system1} (MI~4~RP).}
\label{tab:all_results}
\begin{center}
\setlength{\extrarowheight}{3pt}
\begin{tabular}{c|rr|rr|rr|rr|rr|rr|rr|rr|rr}
\hline
%\noalign{\smallskip}
  & \multicolumn{2}{c|}{LT 6 s} &  \multicolumn{2}{c|}{LT 6 s s} & \multicolumn{2}{c|}{LT s 1} &  \multicolumn{2}{c|}{G 6 s} & \multicolumn{2}{c|}{G s 1} & \multicolumn{2}{c|}{G RP} & \multicolumn{2}{c|}{G RP c} & \multicolumn{2}{c|}{MI 6 RP} &\multicolumn{2}{c}{MI 4 RP} \\
  & Mean & Std & Mean & Std & Mean & Std & Mean & Std & Mean & Std & Mean & Std & Mean & Std & Mean & Std & Mean & Std \\
%\noalign{\smallskip}
\hline
%\noalign{\smallskip}
$g_{1,1}$ & -0.07 &	0.02 &	\com{-0.05} & \com{0.01} & 0.04	& 0.10	& -0.13	   & 0.13  & -0.04	& 0.07	& -0.02	& 0.06	& -0.06	& 0.05	& -0.06	 &   0.02   & 	-0.09	&  0.11     \\
$g_{1,2}$ & -0.45 &	0.04 & \com{-0.47} & \com{0.01} & -0.35	& 0.32	& -0.63	   & 0.11  & -0.55	& 0.05	& -0.56	& 0.06	& -0.58	& 0.08	& -0.61	 &   0.06   & 	-0.87	&  0.35     \\
%\noalign{\smallskip}
\hline
%\noalign{\smallskip}
$\mathcal{J}_{2}$ & -0.36 &	0.08 &\com{-0.35} & \com{0.06} & -0.28	& 0.16	& -0.28	   & 2.34  & -1.38	& 0.34	& -1.30	& 0.31	& -0.74	& 0.33	& -0.42	 &   0.04   & 	-0.36	&  0.08     \\
$f_{2}$ &  2.23 &	0.49 &	\com{2.11} & \com{0.25} & -10.36	& 10.92	& -17.41   & 49.1  & -0.50	& 37.00	& -0.95	& 11.85	& 10.75	& 45.84	& 1.94	 &   0.46   & 	0.42	&  63.49    \\
$g_{2,1}$ &  0.50 & 0.22 &	 \com{0.56} & \com{0.15} & -4.76	& 4.70	& 6.89	   & 17.4  & -0.30	& 12.40	& 0.33	& 3.45	& -2.04	& 10.06	& -0.27	 &   0.10   & 	-0.05	&  14.29    \\
$g_{2,2}$ & -3.42 &	0.55 & \com{-3.32} & \com{0.66} & 15.66	& 16.85	& 14.25	   & 41.8  & 0.44	& 23.64	& 0.48	& 8.16	& -9.34	& 35.58	& -2.55	 &   0.61   & 	-0.08	&  96.97    \\
%\noalign{\smallskip}
\hline
%\noalign{\smallskip} 
$\mathcal{J}_{3}$ &  0.93 &	0.11 &	\com{0.84} & \com{0.03} &  0.66	& 0.20	& 0.52	   & 0.21  & 0.87	& 0.29	& 0.88	& 0.19	& 0.65	& 0.21	& 0.72	 &   0.06   & 	0.51	&  0.25     \\
$f_{3}$ &  0.94 &	0.06 &	\com{0.94} & \com{0.01} & 0.84	& 0.01	& 0.82	   & 0.07  & 0.89	& 0.06	& 0.91	& 0.05	& 0.81	& 0.10	& 0.93	 &   0.09   & 	1.14	&  30.00    \\
$g_{3,1}$ & -0.47 &	0.03 & \com{-0.43} & \com{0.03} & -0.36	& 0.12	& -0.47	   & 0.07  & -0.55	& 0.12	& -0.56	& 0.09	& -0.54	& 0.07	& -0.50	 &   0.04   & 	-0.28	&  16.28    \\
$g_{3,2}$ &  0.09 &	0.04 &\com{ 0.06} & \com{0.01} &  0.22	& 0.09	& 0.15	   & 0.07  & 0.09	& 0.06	& 0.13	& 0.06	& 0.14	& 0.06	& 0.06	 &   0.09   & 	-0.42	&  33.12    \\
%\noalign{\smallskip}
\hline
%\noalign{\smallskip}
$\mathcal{J}_{4}$ & -1.20 &	0.27 & \com{-1.62} & \com{0.44} & -2.28	& 2.07	& -0.54	   & 0.28  & -1.07	& 0.42	& -1.12	& 0.19	& -0.88	& 0.23	& -0.73	 &   0.07   & 	-0.61	&  0.14     \\
$f_{4}$ &  1.78 &	0.09 &	\com{1.59} & \com{0.15} & 2.35	& 1.14	& 1.43	   & 1.82  & 13.36	& 37.97	& 4.02	& 1.61	& 2.56	& 0.74	& 1.68	 &   0.30   & 	1.61	&  7.62     \\
$g_{4,1}$ & -0.06 &	0.05 & \com{-0.02} & \com{0.01} & -0.45	& 0.65	& 0.11	   & 0.65  & -0.05	& 2.80	& 0.18	& 0.50	& 0.11	& 0.24	& 0.21	 &   0.09   & 	0.18	&  2.54     \\
$g_{4,2}$ & -1.78 &	0.15 & \com{-1.52} & \com{0.18} & -3.27	& 2.45	& -1.78	   & 1.77  & -11.00	& 29.26	& -3.25	& 1.03	& -2.45	& 0.58	& -1.83	 &   0.26   & 	-1.85	&  8.68     \\
%\noalign{\smallskip}
\hline
%\noalign{\smallskip}
$\mathcal{J}_{5}$ &  0.23 &	0.11 &	\com{0.32}  & \com{0.04} & 1.34	& 1.83	& 0.36	   & 0.30  & 0.31	& 0.18	& 0.32	& 0.12	& 0.23	& 0.07	& 0.18	 &   0.03   & 	0.12	&  0.07     \\
$f_{5}$ & -0.46 &	0.13 & \com{-0.56} & \com{0.03} & -0.43	& 0.34	& -0.60	   & 0.15  & -0.61	& 0.11	& -0.67	& 0.13	& -0.57	& 0.09	& -0.49	 &   0.05   & 	-0.45	&  6.87     \\
$g_{5,1}$ & -0.15 &	0.04 & \com{-0.18} & \com{0.01} & -0.17	& 0.15	& 0.06	   & 0.26  & -0.09	& 0.22	& 0.00	& 0.10	& 0.03	& 0.09	& 0.08	 &   0.06   & 	0.03	&  8.86     \\
$g_{5,2}$ & -0.39 &	0.11 & \com{-0.45} & \com{0.03} & -0.41	& 0.53	& -0.48	   & 0.41  & -0.03	& 0.22	& -0.13	& 0.16	& -0.16	& 0.15	& -0.41	 &   0.08   & 	-0.87	&  13.72    \\
%\noalign{\smallskip}
\hline                                                          
\end{tabular}
\end{center}
\end{sidewaystable*}

\begin{figure*}[t]
\centering
  \includegraphics[width=0.85\textwidth]{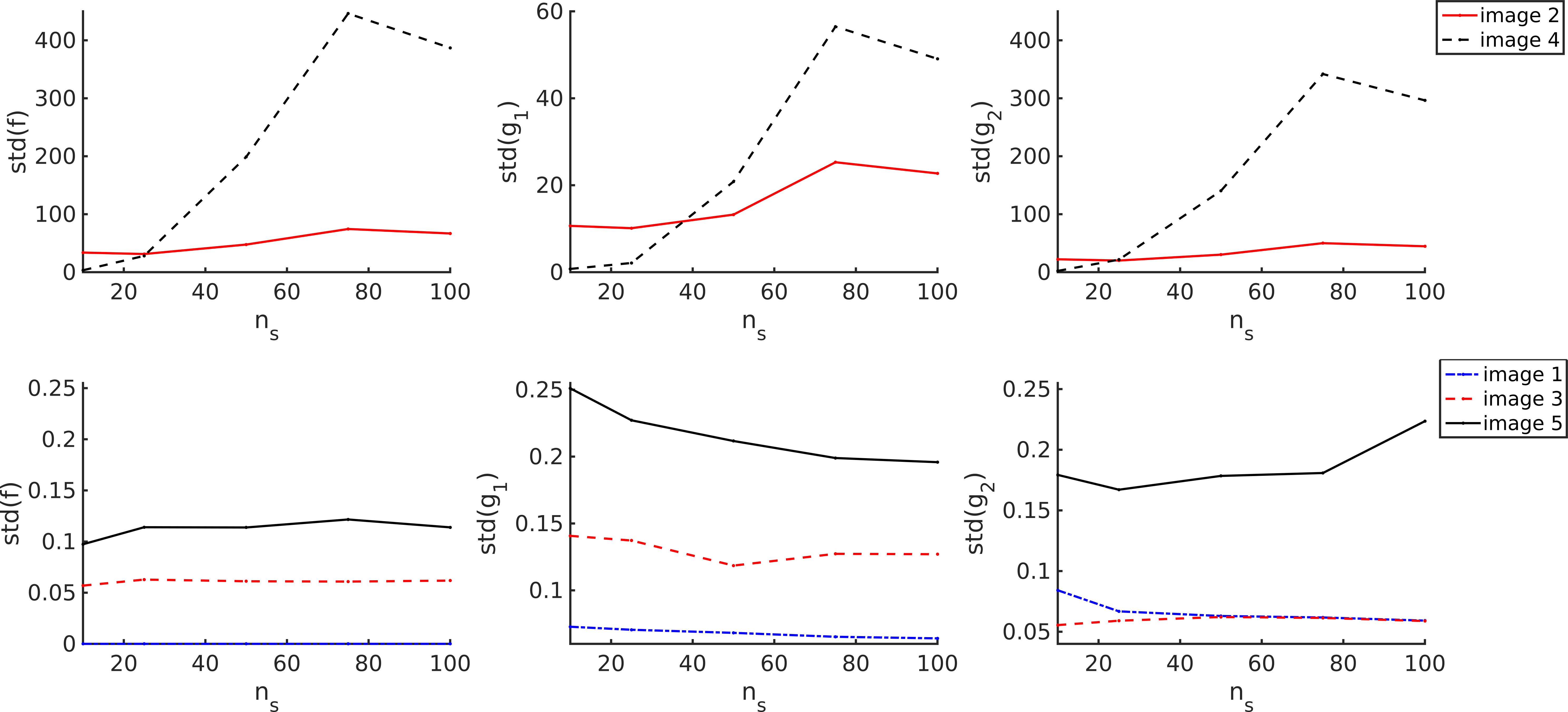}
   \caption{Dependency of the size of the confidence bound, i.e. the standard deviation, of the $f_i,  g_{i,1}$, and $g_{i,2}$, $i=1,...,5$, for all multiple images of system~1 of Table~\ref{tab:multiple_image_systems} determined by the Grale model detailed in Section~\ref{sec:grale_one_image_system}, on the number of individual models $n_\mathrm{s}$ generated by the genetic algorithm and averaged over to obtain the final Grale model.}
\label{fig:grale_statistics}
\end{figure*}

%%%%
\subsection{Comparison of model-independent approaches}
\label{sec:mi_comparison}

Comparing the $\mathcal{J}_i, f_i, g_{i,1}$, and $g_{i,2}$, $i=1,...,5$, using four and six reference points, we find that both reconstructions agree within their standard deviations and the confidence level bounds set by the 16th and 84th percentile \comm{(see Figures~\ref{fig:MI_RP6} and \ref{fig:MI_RP4})}. This implies that the convergence and shear can still be approximated as constant over the area enclosed by the six reference points. The strong disagreement between the mean values for the second image implies steep changes in the convergence and shear fields in the vicinity of the critical curve. Using the four reference points, the larger standard deviations occur because these four reference points cover a smaller area of the image and are more aligned than all six reference points, which increases the uncertainty in the transformation between the reference points that confines the $\mathcal{J}_i, f_i,  g_{i,1}$, and $g_{i,2}$. Evidence supporting this hypothesis can be found in Appendix~\ref{app:mi_confidence_bounds}, in which we calculate the $\mathcal{J}_i, f_i, g_{i,1}$, and  $g_{i,2}$ using four of the six reference points that form a tetragon such that the same area is covered. \com{Hence, the area covered by the reference points and not the number of reference points is decisive for the width of the confidence bounds of the local lens properties.}

Thus, the $\mathcal{J}_i, f_i, g_{i,1}$, and $g_{i,2}$ determined from the six reference points are best suited for the comparison to the model-based approaches. \com{The model-independent approach makes the least amount of assumptions about the lensing configuration. The resulting local lens properties therefore set limits on the precision of the local $\mathcal{J}$-, $f$-, and $g$-values obtainable in a purely data-driven way.} We assume that the true magnification ratios, ratios of convergences, and reduced shear values lie within these confidence bounds, as supported by the simulations of \cite{bib:Wagner2}. \comm{Further accuracy tests using realistically simulated lenses are currently under development.}

%%%%
\subsection{Comparison of Lenstool approaches}
\label{sec:lenstool_comparison}

\com{The three Lenstool models agree in 13 out of 18 $\mathcal{J}_i, f_i,  g_{i,1}$, and $g_{i,2}$, $i=1,...5$ within their confidence bounds. Comparing the first two models using all multiple image systems of Table~\ref{tab:multiple_image_systems}, we find that they agree in all parameters within their confidence bounds. The first model with constant mass-to-light ratio for the cluster member galaxies has mostly larger confidence bounds than the second model.} Using only system~1 of Table~\ref{tab:multiple_image_systems}, we observe larger confidence bounds than when employing all multiple image systems of Table~\ref{tab:multiple_image_systems}.

Given this high degree of agreement, the parametric lens modelling approach yields robust local ratios of convergences and reduced shears, taking into account that the number of constraints from system~1 only suffices to determine the parameters of one large-scale dark matter halo and of the parameters for the smaller-scale dark matter clumps belonging to the brightest member galaxies, while the models employing all six multiple image systems constrain two large-scale dark matter halos and the parameters of the smaller-scale dark matter clumps underneath the brightest member galaxies. In addition, the low number of parameters to be adjusted (8 when using system~1 and 21 \com{(22, when changing the mass-to-light ratio)} when using all six multiple image systems) avoids overfitting to the constraints and the generation of small-scale mass \comm{artefacts}. 

\com{As the difference in the first two Lenstool models is not significant, we compare both of them to the model-independent approach.}

%All $\mathcal{J}_i, f_i,  g_{i,1}$, and $g_{i,2}$, $i=1,...5$, except $f_3$ agree within their confidence bounds when comparing the two Lenstool models in the second and third column block of Table~\ref{tab:all_results} with each other. 

% it does not make sense to compare masses, MI does not provide one, Grale neither. 
% velocity dispersion is another integrated quantity, hard to compare with the measured value from Lenstool parameters...
% only local properties are of interest here!

%%%%
\subsection{Comparison of Grale approaches}
\label{sec:grale_comparison}

% size of the confidence bounds
The Grale models with their many degrees of freedom also have broad confidence bounds. All Grale models yield highly unreliable $\mathcal{J}_i, f_i,  g_{i,1}$, and $g_{i,2}$ for $i=2$ and 4, for which $\kappa \approx 1$. To investigate the reason for the broad confidence bounds, we generate 100 models by the genetic algorithm using only system~1 of Table~\ref{tab:multiple_image_systems} as detailed in Section~\ref{sec:grale_one_image_system}. Averaging over $n_\mathrm{s} = 10, 25, 50, 75, 100$ individual models, five different Grale models are obtained. If the size of the confidence bounds is dominated by statistical uncertainties, we expect the standard deviations to shrink when averaging over an increasing number of individual models. Plotting the standard deviations for these five Grale models for each of the multiple images of system~1 in Figure~\ref{fig:grale_statistics}, we observe that the deviations do not decrease when averaging over an increasing number of individual models. Hence, the size of the confidence bounds is mainly determined by the variation between the different models fulfilling the same (sparse) constraints set by the multiple images. This hypothesis is supported by the fact that the Grale models using all six reference points (see Sections~\ref{sec:grale_all_reference_points} and \ref{sec:grale_all_reference_points_plus_corrections}) yield tighter confidence bounds on the $\mathcal{J}_i, f_i,  g_{i,1}$, and $g_{i,2}$ than the models with only one constraint per multiple image.
\com{Hence, to obtain the tighest confidence bounds on local lens properties, Grale requires as many constraints as possible in the vicinity of the point where the lens properties are to be determined.}

% bias using model-predicted redshifts by Zitrin and comparison to Lenstools LTM
For the comparison shown in Table~\ref{tab:all_results}, we employed the redshifts from \cite{bib:Zitrin} for the Grale model of Section~\ref{sec:grale_six_image_systems} to avoid introducing a bias between the Lenstool and Grale models. \com{After the indepedent lens model comparison}, we \com{now} investigate the influence of the Lenstool redshifts on the Grale model. \com{We employ the redshifts determined by Lenstool in the configuration of Appendix~\ref{app:lenstool_ip_configuration}} (see Section~\ref{sec:lenstool_six_image_systems}) and rerun the model generation procedure detailed in Section~\ref{sec:grale_six_image_systems}. The resulting overall RMSI is $1.16 \pm 0.59$ and thus worse than for the model of Section~\ref{sec:grale_six_image_systems}. As further detailed by the RMSI for the individual multiple image systems and the $\mathcal{J}_i, f_i,  g_{i,1}$, and $g_{i,2}$ in Appendix~\ref{app:grale_redshifts}, the Lenstool redshifts do mainly increase the size of the confidence bounds compared to the Grale model of Section~\ref{sec:grale_six_image_systems} and the large overlap of their confidence bounds does not hint at significant differences between both models. Thus, the redshifts of additional multiple images used for the Grale lens modelling have a minor impact on the $\mathcal{J}_i, f_i,  g_{i,1}$, and $g_{i,2}$ at the positions of system~1 of Table~\ref{tab:multiple_image_systems}.

% physically reasonable mass distribution without overfitting  and discussion of the correction step
As for the other approaches, all Grale models agree within their confidence bounds. This is expected because, as stated in \cite{bib:Ponente}, non-parametric lens modelling approaches can reproduce the multiple images to any accuracy level. Yet, overfitting is prevented in our models by not adding any small-scale mass corrections for the models detailed in Sections~\ref{sec:grale_six_image_systems}, \ref{sec:grale_one_image_system}, and \ref{sec:grale_all_reference_points}. But, as Table~\ref{tab:all_results} shows, even adding small-scale mass corrections as detailed in Section~\ref{sec:grale_all_reference_points_plus_corrections}, does not significantly change the resulting $\mathcal{J}_i, f_i,  g_{i,1}$, and $g_{i,2}$ nor the overall smooth shape of the critical curves (see Figure~\ref{fig:grale_models2}). Thus, we can conclude that \com{local constraints suffice to obtain local lens properties using Grale.} The best Grale model is the one of Section~\ref{sec:grale_all_reference_points_plus_corrections} due to its tightest confidence bounds.

%A comparison with their counterparts from the Grale model without the small-scale mass corrections as shown in the same figure on the left, does not hint at a grave overfitting and unrealistically oscillating mass profiles, which is expected because Grale limits the amount of mass per grid cell in the correction step to suppress this effect.

\begin{figure*}[ht!]
\centering
  \includegraphics[width=0.35\textwidth]{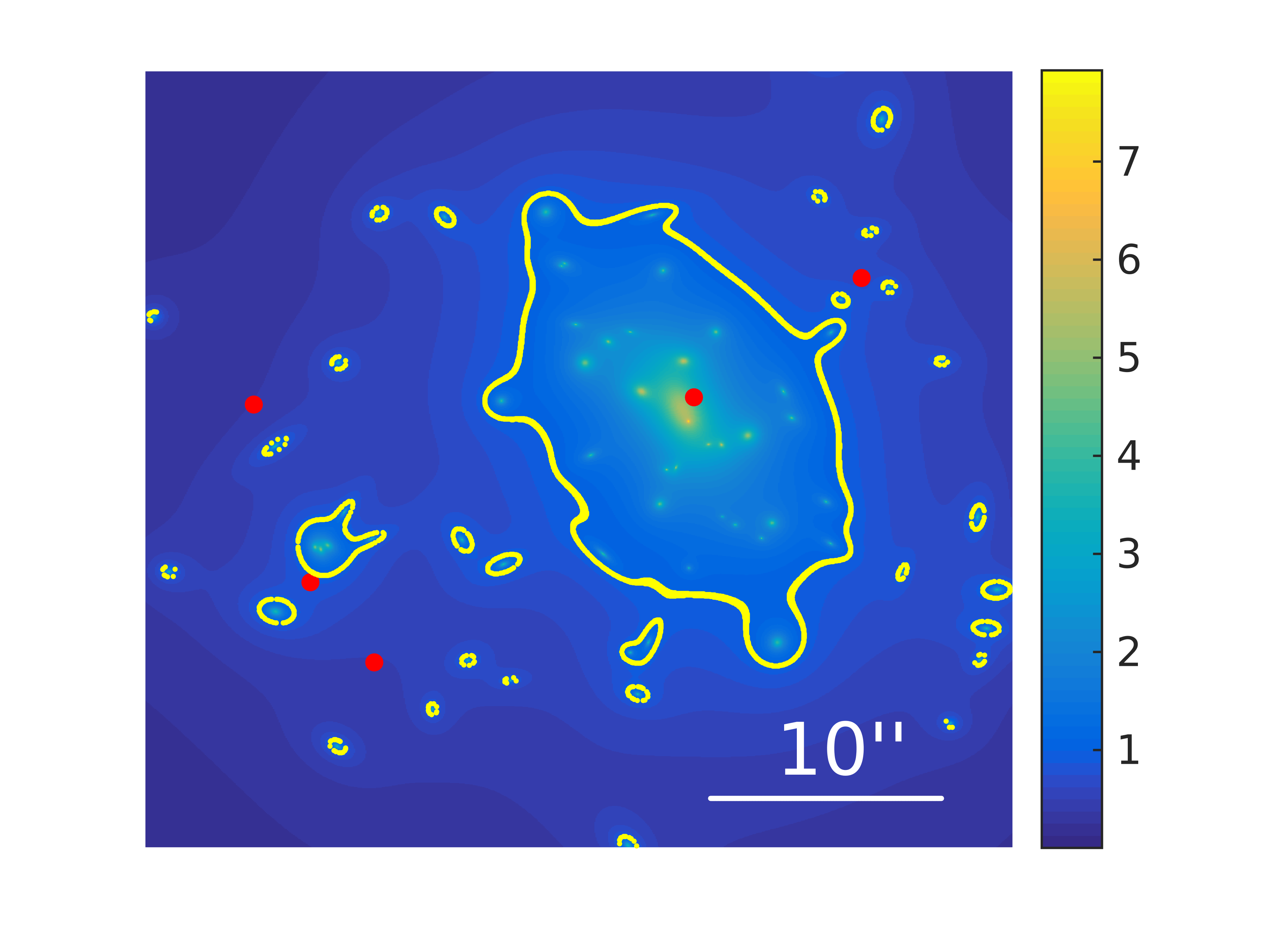}
  \hspace{5ex}
    \includegraphics[width=0.35\textwidth]{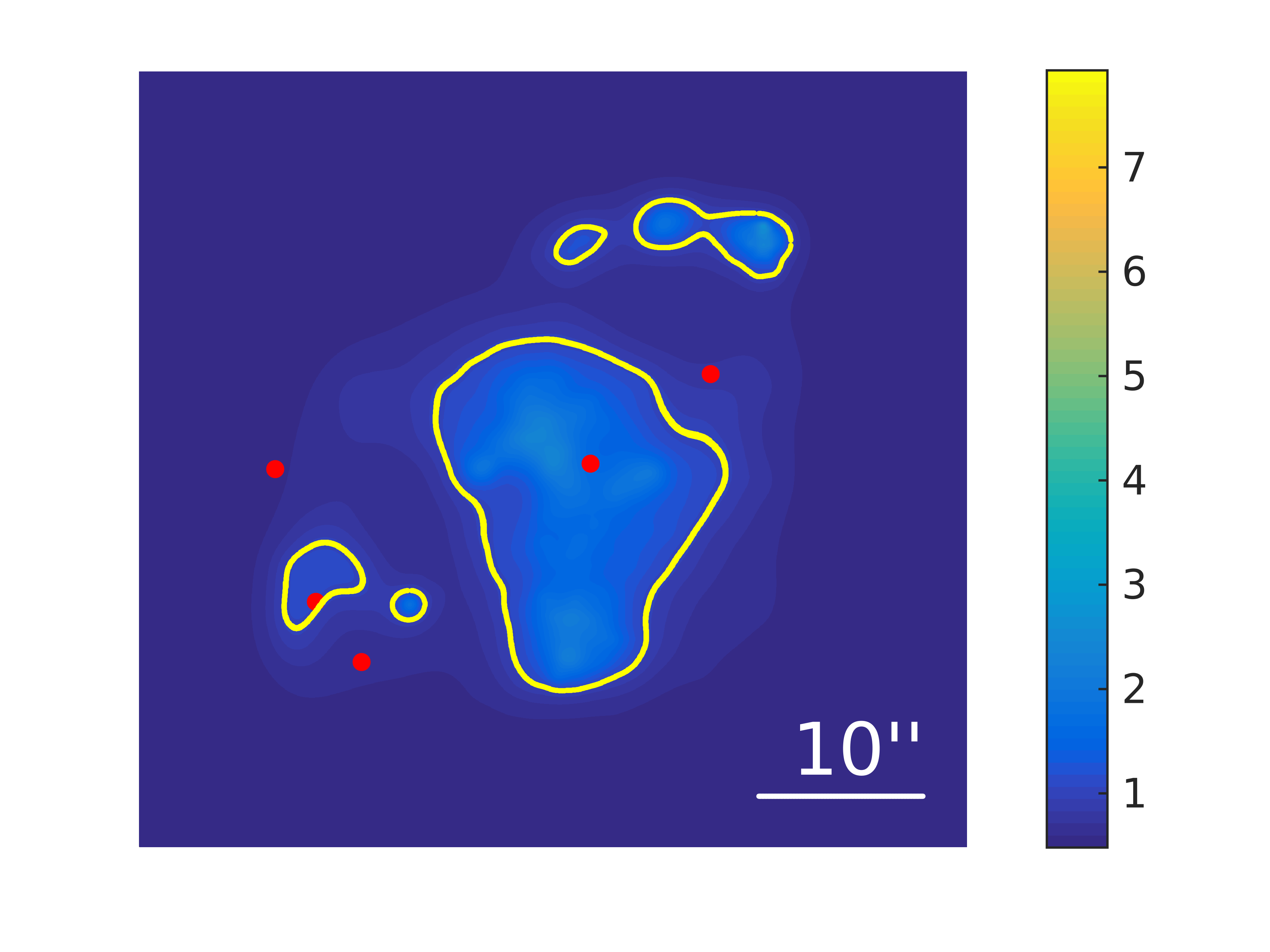}
   \caption{Convergence maps of the Lenstool model of Section~\ref{sec:lenstool_six_image_systems} (left) and the Grale model of Section~\ref{sec:grale_six_image_systems} (right). The positions of system~1 of Table~\ref{tab:multiple_image_systems} are marked in red, the yellow curves delineate the isocontour $\kappa = 1$.}
\label{fig:kappa_maps}
\end{figure*}

%%%%
\subsection{Comparison between all approaches}
\label{sec:comparison_all}

% most robust method (size of confidence bounds)
All approaches show confidence bounds as large as the ratios of convergences and reduced shear values, except for the Lenstool models generated from six multiple image systems, in which the largest confidence bound is 83\% of $g_{4,1}$ \com{for the model with constant mass-to-light ratio of the brightest member galaxies and 50\%  of $g_{4,1}$ for the model with non-constant mass-to-light ratio}. Hence, Lenstool yields the most robust local lens properties, also in the vicinity of the critical curve, comparing, e.g. the confidence bounds of $f_2$ among the approaches. 

% is it sufficient to just use one system as constraints? --> not really
To obtain the $\mathcal{J}_i, f_i,  g_{i,1}$, and $g_{i,2}$, $i=1,...,5$, with the tightest confidence bounds, Lenstool requires several multiple image systems as constraints, while Grale best uses several reference points from a single resolved multiple image system in the vicinity of the point at which the lens properties are to be retrieved. For the model-independent approach, the area over which the reference points are spead is anticorrelated with the size of the confidence bounds. Hence, the area should be maximised, adhering to the approximation that $\mathcal{J}$, $f$ and $g$ are constant.

\com{We thus find that the three methods require complementary constraints from the multiple images and obtain similar local lens properties with comparable precision.}

% proximity to model-independent approach taken as closest to ground truth
Comparing the $\mathcal{J}$-,$f$-, and $g$-values of the model-independent approach  from six reference points to the respective values obtained by the optimum Lenstool models of Section~\ref{sec:lenstool_six_image_systems} and the optimum Grale model of Section~\ref{sec:grale_all_reference_points_plus_corrections}, we find that for Lenstool, 12 \com{(11, for non-constant mass-to-light ratio)} of all 18 $\mathcal{J}$-, $f$-, and $g$-values agree with the model-independently obtained values within their confidence bounds and for Grale, the agreement is found in 17 $\mathcal{J}$-, $f$- and $g$-values. Lenstool deviates in $g_{1,2}, g_{2,1}, \mathcal{J}_{3}, \mathcal{J}_{4}, g_{4,1}$, \com{($\mathcal{J}_5$)}, and $g_{5,1}$, Grale in $g_{5,2}$. 

\com{As the first Lenstool model with constant mass-to-light ratio} agrees to the majority of $\mathcal{J}$-, $f$- and $g$-values of the model-independent approach and only deviates in $g_{1,2}$ and $g_{5,1}$ from the best Grale model, our results are in agreement with the assumption that light traces mass in CL0024. 

\com{The second Lenstool model with non-constant mass-to-light ratio and tighter confidence bounds disagrees in 8 of the 18 parameters with the best Grale model. Hence, we can conclude that a non-constant mass-to-light ratio for the cluster member galaxies is less consistent with the model-independent approach and the best Grale model, so that a constant mass-to-light ratio is favoured in CL0024. The best-fit Lenstool model is thus the one using six multiple image systems and a constant mass-to-light ratio for the brightest member galaxies.}

 %mass-to-light ratio for the small-scale dark matter distribution underneath the brightest galaxies cannot be responsible for the deviations of the Lenstool results because the second model in Table~\ref{tab:all_results} does not show a better agreement to the model-independent approach in the nineth column block.}

% no overfitting in Grale w.r.t to Lenstool results (artifacts from Ponente)
%As Table~\ref{tab:all_qof} shows, the RMSI for the Grale and the Lenstool model using six multiple image systems are comparable. A comparison between the critical curves shown on the left-hand sides of Figures~\ref{fig:lenstool_models} and \ref{fig:grale_models} indicates that the Grale model has not been fine tuned more than Lenstool, as, for instance, the critical curve does not cross the distance between the first and second image of system~10 in both models. Hence, as already discussed in Section~\ref{sec:grale_comparison}, there is no sign of overfitting artifacts as found in \cite{bib:Ponente} in our non-parametric lens models. 

% system 3
Comparing all results gained in Sections~\ref{sec:selection_of_images}, \ref{sec:lenstool_six_image_systems}, \ref{sec:grale_six_image_systems}, and Appendix~\ref{app:grale_redshifts}, the question arises whether system~3 of Table~\ref{tab:multiple_image_systems} is a real multiple image system because its RMSI is higher than the ones of the other multiple image systems in most of the lens models, and the redshift predicted by \com{the best-fit} Lenstool \com{model} is much higher than the one obtained by \cite{bib:Zitrin} and their photometric redshift estimates. \com{Furthermore, two authors independently arrive at the result that system~3 is hard to model with Grale and Lenstool.} Hence, spectroscopic measurements are required to corroborate \com{or reject} the lensing hypothesis for this multiple image system.

% merger hypothesis
Finally, we use the results obtained in Sections~\ref{sec:lenstool_six_image_systems} and \ref{sec:grale_six_image_systems} to investigate the merger hypothesis for CL0024. In the Lenstool and Grale models discussed in these sections, we find deviations from a symmetric, relaxed cluster structure, \comm{as can be observed in the convergence maps of the models from Sections~\ref{sec:lenstool_six_image_systems} and \ref{sec:grale_six_image_systems} shown in Figure~\ref{fig:kappa_maps}: Inspecting the isocontour $\kappa = 1$ (yellow curves) in both convergence maps, the asymmetric shape is clearly seen. While the convergence map reconstructed by Grale shows only few closed curves of $\kappa=1$, several more are observed in the convergence map obtained by Lenstool due to the dark matter halos of the brightest cluster member galaxies. Both convergence maps show regions of $\kappa=1$ close to image~2, as expected from the relatively broad confidence bounds for this image and tend to a similar streching around the central image~5. The differences in the shape of the isocontour for the convergence maps of Lenstool and Grale once again (see also Figure~\ref{fig:grale_statistics}) show the freedom lens models have to extend the lens reconstruction beyond the vicinity of multiple images.} 

The estimation of redshifts is degenerate with the parameters of the dark matter halos in Lenstool. Thus, \comm{beyond a corroboration of the merger hypothesis,} more quantitative statements about the merger masses and geometry cannot be made without spectroscopic redshift measurements.

%%%%%%%%%%%%%%%%%
\section{Conclusion}
\label{sec:conclusion}

We performed the most direct comparison between the model-independent local lens reconstruction approach for multiple images with resolved brightness features as described in \cite{bib:Tessore1, bib:Wagner2}, the parametric lens modelling software Lenstool, \cite{bib:Kneib, bib:Jullo2007}, and the non-parametric lens modelling approach Grale, \cite{bib:Liesenborgs2010}: Using the same positions of multiple images, the same cosmological parameter values, and the same number of model-predicted convergence and shear maps for the evaluation statistics, we determined magnification ratios, ratios of convergences and reduced shears at the positions of the five multiple images of the source at redshift $z_\mathrm{s} = 1.675$ in the galaxy cluster CL0024, \cite{bib:Colley} from both lens modelling approaches and compared these local lens properties to their model-independent counterparts. 

Summarising the results detailed in Section~\ref{sec:comparison}, we arrive at the following conclusions:
\begin{itemize}
\item The local lens properties, i.e.\ the magnification ratios, ratios of convergences, and reduced shear values ($\mathcal{J}$-, $f$-, and $g$-values) at the five positions obtained by the model-independent approach, Lenstool, and Grale coincide in the majority of cases within their confidence bounds, supporting the validity of the light-traces-mass assumption in CL0024 \com{and favouring a constant mass-to-light ratio for the brightest cluster member galaxies.}
\item Our results are in agreement with the merger hypothesis assumed in \cite{bib:Kneib, bib:Zhang, bib:Zitrin} because, according to our Lenstool models, the smallest, most probable number of large-scale dark matter halos for the strong lensing region is two and the convergence maps generated by Grale also suggest perturbations to a symmetric, relaxed cluster structure (see \com{Figure}~\ref{fig:kappa_maps}). 
\item Our Lenstool and Grale models mostly encountered high root-mean square deviations between the observed and model-predicted positions of the multiple images of system~3 in Table~\ref{tab:multiple_image_systems} compared to all other multiple image systems employed. \com{The best-fit} Lenstool \com{model} also predicted a much higher redshift for this system ($3.49 \pm 0.39$) than \cite{bib:Zitrin} ($2.55^{+ 0.45}_{- 0.20}$) and the photometric redshift estimates (between $2.48$ and $2.76$, see \cite{bib:Zitrin}). Hence, spectroscopic observations are necessary to further investigate whether these observations really originate from the same source galaxy. 
\item All three approaches show broad confidence bounds for the $f$-, and $g$-values that can become as large as the values themselves, especially close to regions where the convergence equals one and the denominator in the $f$s and $g$s approaches zero. This is in agreement with the findings made by \cite{bib:Meneghetti2016} employing unresolved multiple image systems. 
\item From a methodological point of view, we discovered that the model-independent approach yields $\mathcal{J}$-, $f$-, and $g$-values that are of the same quality as the model-generated ones, if there are at least four resolved brightness features forming a tetragon that covers an image region of approximately constant convergence and shear. While Lenstool is well-suited to reconstruct the global cluster structure including the member galaxies, local lens properties on (sub-)galaxy scale are better determined by Grale when reconstructing the local $f$- and $g$-values from all resolved brightness features close to the position of interest and adding limited small-scale mass corrections. This is very advantageous because, in this way, no unconfirmed additional multiple image systems with uncertain photometric redshifts have to be taken into account. Limiting the mass corrections to 10\% of the total mass, Grale shows no sign of unrealistically oscillating mass distributions. 
\item For the run times, we find that the model-independent approach takes about 0.23 s to determine the values in the last two column blocks of Table~\ref{tab:all_results} using a Linux-PC with 8 $\times$ Intel Core i7-4710MQ CPU @ 2.50GHz and 31.1~GiB RAM. On the same machine, Lenstool, Version 6.8.1., takes about 24~h for each of the 40 models of Section~\ref{sec:lenstool_six_image_systems} and ca.\ 4~h for each of the 40 models of Section~\ref{sec:lenstool_one_image_system} including the calculations for the convergence and shear maps. The Grale algorithm takes about 45~min to obtain one individual model of the genetic algorithm for the specifications of Section~\ref{sec:grale_six_image_systems}, 10~min for one individual model for Sections~\ref{sec:grale_one_image_system} and \ref{sec:grale_all_reference_points}, and 30~min to determine the small-scale mass corrections for one individual model in Section~\ref{sec:grale_all_reference_points_plus_corrections}, running on a single computing node with 2 $\times$ 12-core "Haswell" processors of type Xeon E5-2680v3. \com{Thus, the model-independent approach not only employs the miminum set of assumptions about the lensing configuration but is by far the fastest way to extract the local lens properties, as well.}
\end{itemize}

% LTM support also by Diego et al 2005? 
% \item do the results agree with the results obtained in simulations as discussed in \cite{bib:Meneghetti2016}? 

%%%%%%%%%%%%%%%%%
\begin{acknowledgements}
JW would like to thank Mauricio Carrasco, Matteo Maturi, Massimo Meneghetti, Sven Meyer, Juan Remolina, Sebastian Stapelberg, R\"udiger Vaas, and Adi Zitrin for helpful discussions, and Johan Richard for helpful discussions and providing the information on his lens model. JW gratefully acknowledges the support by the Deutsche Forschungsgemeinschaft (DFG) WA3547/1-1. JL acknowledges the use of the computational resources and services provided by the VSC (Flemish Supercomputer Center), funded by the Research Foundation - Flanders (FWO) and the Flemish Government – department EWI. NT acknowledges support from the European Research Council in the form of a Consolidator Grant with number 681431.
\end{acknowledgements}

\bibliographystyle{aa}
\bibliography{aa}

%%%%%%%%%%%%%%%%%%%%%%%%%%%%%%%%%%%%%%%%%%%%%%%%%%%%%%%%%%%%%%%%%%%%%%%%%%%%%
\appendix

\section{Lenstool configuration file for the model in Section~\ref{sec:selection_of_images}}
\label{app:lenstool_sp_configuration}
\begin{verbatim}
runmode
	  reference  3 6.648555 17.161900
	  inverse    3 0.3 10
	  image      3 mult_images_zitrin_deg12.cat
	  mass       1 12000 0.39 convergence.fits
	  shear      3 12000 1.675 gamma1.fits
	  shear      4 12000 1.675 gamma2.fits
	  end
image
	  multfile  1  mult_images_zitrin_deg12.cat
	  mult_wcs  1
	  sigposArcsec 0.2
	  z_m_limit 1 P03 1 0.39 5.0 0.1
	  z_m_limit 1 P04 1 0.39 5.0 0.1
	  z_m_limit 1 P05 1 0.39 5.0 0.1
	  z_m_limit 1 P08 1 0.39 5.0 0.1
	  z_m_limit 1 P10 1 0.39 5.0 0.1
	  forme      0
	  end
grille
	  nombre           128
	  polaire          0
	  nlentille        86
	  nlens_opt        1
	  end
source 
  z_source         1.675
  end
potentiel 1000
	  profil           81
	  x_centre         0.0
	  y_centre         0.0
	  ellipticite	 	   0.107
	  angle_pos 	      80.60
	  core_radius_kpc 	2.023
	  cut_radius_kpc  	1000.00
	  v_disp           1201.986
	  z_lens           0.3900
	  end
limit 1000
  x_centre         1 -50.0   50.00 0.20
  y_centre         1 -50.0   50.00 0.20
  ellipticite      1   0.0    0.95 0.01
  angle_pos        1   0.0  180.00 0.10
  core_radius_kpc  1   0.1  500.00 0.10
  cut_radius       1   5.0 2000.00 0.10
  v_disp           1 100.0 2000.00 1.00
  end
potfile
  filein           1 galsort.cat
  zlens            0.39
  type             81
  x_centre         1 -100.0 100.0 0.05
  y_centre         1 -100.0 100.0 0.05
  corekpc          1    0.1   3.0 0.10
  mag0             20.500000
  sigma            1    0.0 450.0 0.10
  cutkpc           1    0.0 500.000.10
  end
cline
  nplan            1 1.675
  algorithm        MARCHINGSQUARES
  limitHigh        10.0
  limitLow         3.0
  end
cosmologie
  H0                67.800
  omegaM            0.308
  omegaX            0.692
  omegaK            0.000
  wX               -1.000
  end
champ
  xmin             -300
  xmax              300
  ymin             -300
  ymax              300
  end
fini
\end{verbatim}

\setcounter{savecounter}{\value{section}} % Save the current section
\setcounter{section}{3} % Set the current section to 2 (appendix B) 
\begin{figure*}[ht]
\centering
  \includegraphics[width=0.32\textwidth]{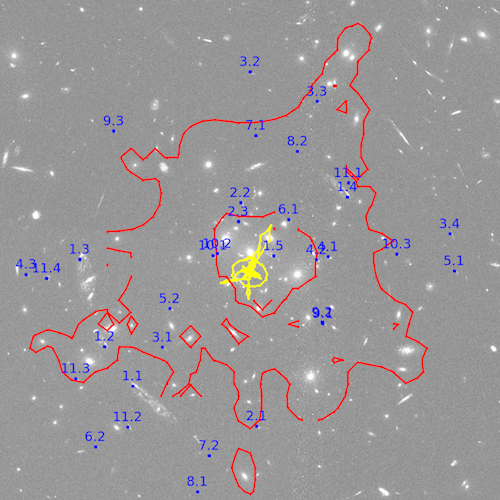}
  \hspace{1ex}
  \includegraphics[width=0.32\textwidth]{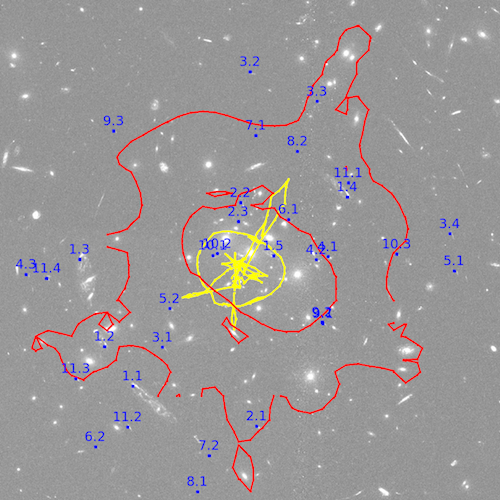}
  \hspace{1ex}
   \includegraphics[width=0.32\textwidth]{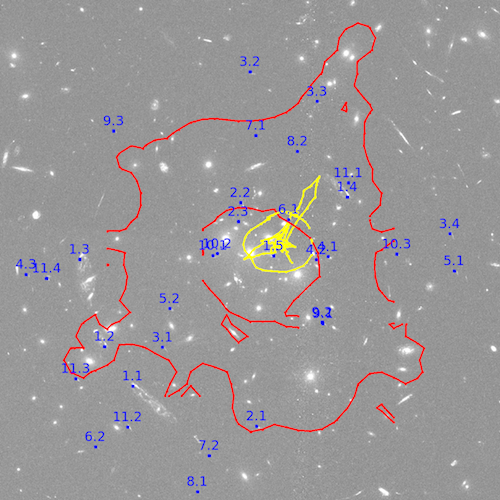}
   \caption{Lenstool lens models employing the 85 brightest member galaxies, system~1 of Table~\ref{tab:multiple_image_systems} and systems~2--11 of \cite{bib:Zitrin}, and one large-scale PIEMD dark matter halo (left), and two large-scale PIEMD dark matter halos (centre), and three large-scale PIEMD dark matter halos (right). The critical curves determined by the marching squares algorithm (see Appendix~\ref{app:lenstool_sp_configuration} for its configuration) are marked in red, the caustics in yellow and the multiple images in the notation of \cite{bib:Zitrin} in blue.}
\label{fig:models}
\end{figure*}
\setcounter{section}{\value{savecounter}} % Set the current section to 2 (appendix B) 

%%%%%%%%%%%%%%%%%
\section{Lenstool configuration file for the model in Section~\ref{sec:lenstool_six_image_systems}}
\label{app:lenstool_ip_configuration}
\begin{verbatim}
runmode
	  reference  3 6.648555 17.161900
	  inverse    3 0.1 100
	  image      3 mult_images_zitrin_deg12.cat
	  mass       1 12000 0.39 convergence.fits
	  shear      3 12000 1.675 gamma1.fits
	  shear      4 12000 1.675 gamma2.fits
	  end
image
	  multfile  1  mult_images_zitrin_deg12.cat
	  mult_wcs  1
	  sigposArcsec 0.2
	  z_m_limit 1 P03 1 0.39 5.0 0.1
	  z_m_limit 1 P04 1 0.39 5.0 0.1
	  z_m_limit 1 P05 1 0.39 5.0 0.1
	  z_m_limit 1 P08 1 0.39 5.0 0.1
	  z_m_limit 1 P10 1 0.39 5.0 0.1
	  forme     -1
	  end
grille
	  nombre           128
	  polaire          0
	  nlentille        86
	  nlens_opt        2
	  end
source 
  z_source         1.675
  end
potentiel 1000
	  profil           81
	  x_centre         0.0
	  y_centre         0.0
	  ellipticite	 	   0.107
	  angle_pos 	      80.60
	  core_radius_kpc 	2.023
	  cut_radius_kpc  	1000.00
	  v_disp           1201.986
	  z_lens           0.3900
	  end
limit 1000
  x_centre         1 -10.0   10.00 0.20
  y_centre         1 -10.0   10.00 0.20
  ellipticite      1   0.0    0.95 0.01
  angle_pos        1   0.0  180.00 0.10
  core_radius_kpc  1   0.1  500.00 0.10
  cut_radius       1   5.0 2000.00 0.10
  v_disp           1 100.0 2000.00 1.00
  end
potentiel 1001
	  profil           81
	  x_centre         0.0
	  y_centre         0.0
	  ellipticite      0.107
	  angle_pos        80.60
	  core_radius_kpc  2.023
	  cut_radius_kpc   1000.00
	  v_disp           201.986
	  z_lens           0.3900
	  end
limit 1001
  x_centre         1 -50.0   50.00 0.20
  y_centre         1 -50.0   50.00 0.20
  ellipticite      1   0.0    0.95 0.01
  angle_pos        1   0.0  180.00 0.10
  core_radius_kpc  1   0.1  500.00 0.10
  cut_radius       1   0.5 2000.00 0.10
  v_disp           1  10   2000.00 1.00
  end
potfile
  filein           1 galsort.cat
  zlens            0.39
  type             81
  x_centre         1 -100.0 100.0 0.05
  y_centre         1 -100.0 100.0 0.05
  corekpc          1    0.1   3.0 0.10
  mag0             20.500000
  sigma            1    0.0 450.0 0.10
  cutkpc           1    0.0 500.000.10
  end
cline
  nplan            1 1.675
  algorithm        MARCHINGSQUARES
  limitHigh        10.0
  limitLow         3.0
  end
cosmologie
  H0                67.800
  omegaM            0.308
  omegaX            0.692
  omegaK            0.000
  wX               -1.000
  end
champ
  xmin             -300
  xmax              300
  ymin             -300
  ymax              300
  end
fini
\end{verbatim}

%%%%%%%%%%%%%%%%%
\section{Lenstool models used in Section~\ref{sec:selection_of_images}}
\label{app:selection}
Using the configuration file of Section~\ref{app:lenstool_sp_configuration} with one PIEMD dark matter halo, system~1 of Table~\ref{tab:multiple_image_systems} and systems~2--11 of \cite{bib:Zitrin}, and the catalogue of the brightest cluster member galaxies, we arrive at a lens model whose critical curves and caustics are shown in Figure~\ref{fig:models}~(left). Adapting the configuration file to two and three PIEMD dark matter halos for the same remaining specifications, we obtain the critical curves and caustics of Figure~\ref{fig:models}~(centre) and \ref{fig:models}~(right), respectively. 

For all models, the results for the goodness-of-fit measures and the degrees of freedom, i.e.\ the number of constraints minus the number of lens model parameters, are summarised in Table~\ref{tab:selection1}.

\begin{table}[t]
\caption{Degrees of freedom (DOF), the logarithm of the evidence ($\log(\mathcal{E})$), root-mean-square deviations in the image plane (RMSI) in arcseconds over all image systems, and total $\chi^2$ of Lenstool models for CL0024 for varying numbers of dark matter halo PIEMDs (\# PIEMDs) using system~1 in Table~\ref{tab:multiple_image_systems} and systems~2--11 of \cite{bib:Zitrin} as constraints.}
\label{tab:selection1}
\begin{center}
\begin{tabular}{ccccc}
\hline
\noalign{\smallskip}
\# PIEMDs & DOF & $\log(\mathcal{E})$ & RMSI & $\chi^2$ \\
\noalign{\smallskip}
\hline
\noalign{\smallskip}
1 & 25 &  -858 & 2.58 & 2025 \\
2 & 18 & -447 & 1.73 & 1105 \\
3 & 11 & -332 & 1.92 & 901 \\
\noalign{\smallskip}
\hline
\end{tabular}
\end{center}
\end{table}

\begin{table}[t]
\caption{Individual RMSI in arcseconds for system~1 in Table~\ref{tab:multiple_image_systems} and systems~2--11 of \cite{bib:Zitrin} in the lens models with one, two, and three PIEMD dark matter halos. For multiple image sytems having RMSI~=~0.00 the barycentre of their back-traced images is not found at the required precision, i.e.\ the lens model might not be able to explain those systems.}
\label{tab:selection2}
\begin{center}
\begin{tabular}{cccc}
\hline
\noalign{\smallskip}
System & RMSI & RMSI & RMSI\\
 & 1 PIEMD & 2 PIEMDs & 3 PIEMDs \\
\noalign{\smallskip}
\hline
\noalign{\smallskip}
 1   &    2.63   &    1.36    &    1.17 \\
 2   &    3.51   &    1.80    &    1.34 \\
 3   &    5.08   &    3.14    &    2.11 \\
 4   &    1.09   &    1.10    &    0.79 \\
 5   &    1.86   &    0.79    &    1.18 \\
 6   &    0.00   &    0.00    &    0.00 \\
 7   &    0.00   &    0.00    &    3.26 \\ 
 8   &    0.99   &    0.35    &    1.20 \\
 9   &    3.06   &    2.28    &    2.85 \\
10   &    0.98   &    1.47    &    0.00 \\ 
11   &    3.46   &    2.94    &    3.29 \\

\noalign{\smallskip}
\hline
\end{tabular}
\end{center}
\end{table}

With the same three lens models considered in Table~\ref{tab:selection1}, we select the set of multiple image systems to generate the best-fit Lenstool model by considering the RMSI for the single multiple image sytems, as shown in Table~\ref{tab:selection2}. 
Systems 6 and 7 are eliminated from our set, as they are  no real multiple image sytems or require further smaller-scale substructure fine-tuning because at least two lens models cannot determine their source positions within the required precision. We also eliminate systems 9 and 11, due to their non-decreasing, high RMSI-values. As the central part of the cluster is already probed by systems~4 and 10 and the remaining image of system~2 is far from all images of system~1, it is also discarded from the set. Although system~3 shows high RMSI, it is kept in the set, as its RMSI decreases quickly with increasing number of PIEMDs and with its far-spread images 3.2, 3.3, and 3.4, it constrains the lensing potential around image 1.4. Thus, the set of multiple image systems as shown in Table~\ref{tab:multiple_image_systems} is obtained.

%%%%%%%%%%%%%%%%%%
%\section{Likelihood distribution for the Lenstool model of Section~\ref{sec:lenstool_six_image_systems}}
%\label{app:LT_6s_histogram}
%
%Figure~\ref{fig:LT_6s} shows the likelihood distributions for the $\mathcal{J}$s, $f$s, and $g$s of the Lenstool model of Section~\ref{sec:lenstool_six_image_systems}. 
%
%
%%%%%%%%%%%%%%%%%%
%\section{Likelihood distribution for the Lenstool model of Section~\ref{sec:lenstool_one_image_system}}
%\label{app:LT_1s_histogram}
%
%Figure~\ref{fig:LT_1s} shows the likelihood distributions for the $\mathcal{J}$s, $f$s, and $g$s of the Lenstool model of Section~\ref{sec:lenstool_one_image_system}. 

%%%%%%%%%%%%%%%%%
\section{Grale configurations for the models used in Section~\ref{sec:Grale}}
\label{app:grale_configuration} 

We employ a 60'' squared region around the reference point in the Lenstool configuration files (see Appendix~\ref{app:lenstool_ip_configuration}). As initial grid, 15 times 15 uniformly distributed squared grid cells are generated and the grid is refined to the level the number and positions of the constraints permit. The prediction of images relatively far from the cluster centre is avoided by introducing a nullspace grid of 200'' edge length centred at the same reference point. 

For the reconstruction using six multiple image systems, about 600 basis functions are used, while for the remaining Grale models, about 300 basis functions are taken into account.

%%%%%%%%%%%%%%%%%
\section{Influence of the area spanned by the reference points on the confidence bounds in the model-independent approach}
\label{app:mi_confidence_bounds}

Instead of discarding the first two reference points as done in Section~\ref{sec:mi_information}, we now discard point~3 and point~5 in Table~\ref{tab:system1} to obtain the following mean and median $\mathcal{J}$-, $f$-, and $g$-values, their confidence intervals set by the 16th and 84th percentile and their standard deviations:
\begin{center}
\setlength{\extrarowheight}{3pt}
\begin{tabular}{c|cc|ccc}
 & Mean & Std & Median & Upper & Lower \\
 &             &         &               & bound & bound\\
% \noalign{\smallskip}
 \hline
% \noalign{\smallskip}
$g_{1,1}$ &$-0.05$ & $0.03$ & $-0.05$ & $0.02$ & $0.03$ \\
$g_{1,2}$ &$-0.59$ & $0.06$ & $-0.60$ & $0.06$ & $0.05$ \\
%\noalign{\smallskip}
\hline
%\noalign{\smallskip}
$\mathcal{J}_2$ & $-0.43$ & $0.04$ & $-0.43$ & $0.04$ & $0.04$ \\
$f_{2}$ & $\phantom{-}2.15$ & $0.80$ & $\phantom{-}2.01$ & $0.57$ & $0.36$\\ 
$g_{2,1}$ & $-0.31$ & $0.13$ & $-0.30$ & $0.11$ & $0.12$ \\
$g_{2,2}$ &  $-2.83$ & $1.11$ & $-2.61$ & $0.44$ & $0.77$ \\
%\noalign{\smallskip}
\hline
%\noalign{\smallskip}
$\mathcal{J}_3$ & $\phantom{-}0.72$ & $0.06$ & $\phantom{-}0.72$ & $0.05$ & $0.05$ \\
$f_{3}$ & $\phantom{-}0.90$ & $0.09$ & $\phantom{-}0.90$ & $0.09$ & $0.09$ \\
$g_{3,1}$ &$-0.51$ & $0.04$ & $-0.51$ & $0.04$ & $0.04$  \\
$g_{3,2}$ & $\phantom{-}0.09$ & $0.09$ & $\phantom{-}0.10$ & $0.09$ & $0.09$  \\
%\noalign{\smallskip}
\hline
%\noalign{\smallskip}
$\mathcal{J}_4$ & $-0.74$ & $0.06$ & $-0.74$ & $0.07$ & $0.06$  \\
$f_{4}$ &  $\phantom{-}1.80$ & $0.34$ & $\phantom{-}1.75$ & $0.35$ & $0.26$ \\
$g_{4,1}$ & $\phantom{-}0.28$ & $0.10$ & $\phantom{-}0.27$ & $0.11$ & $0.10$  \\
$g_{4,2}$ & $-1.94$ & $0.31$ & $-1.89$ & $0.22$ & $0.31$  \\
%\noalign{\smallskip}
\hline
%\noalign{\smallskip}
$\mathcal{J}_5$ & $\phantom{-}0.18$ & $0.03$ & $\phantom{-}0.18$ & $0.03$ & $0.03$  \\
$f_{5}$ & $-0.49$ & $0.05$ & $-0.49$ & $0.05$ & $0.05$  \\
$g_{5,1}$ & $\phantom{-}0.07$ & $0.08$ & $\phantom{-}0.07$ & $0.08$ & $0.08$  \\
$g_{5,2}$ & $-0.38$ & $0.09$ & $-0.38$ & $0.09$ & $0.09$ \\
%\noalign{\smallskip}
\hline
\end{tabular}
\end{center}
Comparing the size of the confidence intervals with those using six reference points (eighth column block of Table~\ref{tab:all_results}, we see that they are of comparable size and smaller than the ones using four reference points spanning a smaller area (last column block of Table~\ref{tab:all_results}).

%%%%%%%%%%%%%%%%%
\section{Influence of redshifts in the Grale model of Section~\ref{sec:grale_six_image_systems}}
\label{app:grale_redshifts}

Instead of employing the model-predicted redshifts of \cite{bib:Zitrin}, we now use the model-predicted redshifts as determined by Lenstool in Section~\ref{sec:lenstool_six_image_systems} to generate the Grale model detailed in Section~\ref{sec:grale_six_image_systems}. The RMSI per multiple image system are as follows, for comparison, we list the RMSI for the Grale model of Section~\ref{sec:grale_six_image_systems} in the last column:
\begin{center}
\begin{tabular}{ccc}
\hline
\noalign{\smallskip}
System & RMSI & RMSI \\
 & (LT $z$) & (Section~\ref{sec:grale_six_image_systems}) \\
\noalign{\smallskip}
\hline
\noalign{\smallskip}
  1  &  $0.91 \pm 0.36$ & $0.68 \pm 0.28$ \\ 
  3 &  $2.05 \pm 1.46$  & $1.52 \pm 2.26$ \\
  4 &   $0.22 \pm 0.38$ & $0.09 \pm 0.18$ \\
  5 &  $0.11 \pm 0.14$  & $0.04 \pm 0.04$ \\  
  8 &  $0.07 \pm 0.15$  & $0.02 \pm 0.02$ \\
10 &  $0.24 \pm 0.65$  & $0.41 \pm 0.70$ \\
\noalign{\smallskip}
\hline
\end{tabular}
\end{center}
From the convergence and shear maps, we obtain the following mean $\mathcal{J}$-, $f$-, and $g$-values and their standard deviations, for comparison, we add the respective values for the model of Section~\ref{sec:grale_six_image_systems} in the fourth and fifth column:
\begin{center}
\setlength{\extrarowheight}{3pt}
\begin{tabular}{c|rr|rr}
 & Mean &  Std & Mean & Std \\
 & \multicolumn{2}{c|}{(LT $z$)} &  \multicolumn{2}{c}{(Sec.~\ref{sec:grale_six_image_systems})} \\
% \noalign{\smallskip}
 \hline
% \noalign{\smallskip}
$g_{1,1}$ & $-0.26$ & $0.23$                        & $-0.13$  & $0.13$ \\
$g_{1,2}$ & $-0.53$ &  $0.12$ & $-0.63$ & $0.11$ \\
%\noalign{\smallskip}
\hline
%\noalign{\smallskip}
$\mathcal{J}_2$ & $-0.78$ & $0.97$                          & $-0.28$ & $2.34$ \\
$f_{2}$                  & $4.62$ & $51.58$ & $-17.41$ &  $49.18$ \\
$g_{2,1}$             & $0.14$ & $21.15$ & $6.89$ &  $17.48$ \\
$g_{2,2}$             & $-5.52 $ & $53.06$ & $14.25$ &  $41.82$ \\
%\noalign{\smallskip}
\hline
%\noalign{\smallskip}
$\mathcal{J}_3$ & $0.51$ & $0.23$  & $0.52$ & $0.21$ \\
$f_{3}$ & $0.80$ & $0.11$                   & $0.82$ & $0.07$ \\
$g_{3,1}$ & $-0.39$ & $0.08$                                     & $-0.47$ & $0.07$ \\
$g_{3,2}$ & $0.23$ & $0.09$                                     & $0.15$ & $0.07$ \\
%\noalign{\smallskip}
\hline
%\noalign{\smallskip}
$\mathcal{J}_4$ &  $-0.64$ & $0.45$          & $-0.54$ & $0.28$ \\
$f_{4}$ & $1.58$ & $2.17$     & $1.43$ & $1.82$ \\
$g_{4,1}$ & $-0.33$ &  $0.74$                        & $0.11$ & $0.65$ \\
$g_{4,2}$ & $-1.90$ & $2.36$ & $-1.78$ & $1.77$ \\
%\noalign{\smallskip}
\hline
%\noalign{\smallskip}
$\mathcal{J}_5$ & $0.28$ & $0.16$   & $0.36$ & $0.30$ \\
$f_{5}$ & $-0.57$ & $0.15$                                           & $-0.59$ & $0.15$ \\
$g_{5,1}$ & $0.21$ & $0.22$               & $0.06$ & $0.26$ \\
$g_{5,2}$ & $-0.28$ & $0.30$               & $-0.48$ & $0.41$ \\
%\noalign{\smallskip}
\hline
\end{tabular}
\end{center}

\end{document}